\documentclass[iop,numberedappendix,twocolappendix,twocolumn]{emulateapj} 

\usepackage{latexsym}   %Simbolos de latex
\usepackage{amsfonts}   %Fuentes de la AMS
\usepackage{amstext}    %Texto de la AMS
\usepackage{amsmath,amssymb}
\usepackage{mathtools}
\usepackage{bm}
\usepackage{MnSymbol} 
\usepackage{mathrsfs}
\usepackage{xcolor}
\usepackage{algorithm2e}
\usepackage{hyperref}
\hypersetup{
  colorlinks   = true, %Colours links instead of ugly boxes
  urlcolor     = blue, %Colour for external hyperlinks
  linkcolor    = blue, %Colour of internal links
  citecolor   = blue %Colour of citations
}

\begin{document}
 
\title{\bf Formal Solutions for Polarized Radiative Transfer\\ IV. Numerical Performances in Practical Problems}
\author{Gioele Janett\altaffilmark{1,2}, Oskar Steiner\altaffilmark{1,3}, Luca Belluzzi\altaffilmark{1,3}}% Edgar Carlin\altaffilmark{1}
\email{gioele.janett@irsol.ch}%, steiner@kis.uni-freiburg.de, belluzzi@irsol.ch, ecarlin@irsol.ch

\affil{$^1$ Istituto Ricerche Solari Locarno (IRSOL), 6605 Locarno-Monti, Switzerland\\
$^2$ Seminar for Applied Mathematics (SAM), ETH Zurich, 8093 Zurich, Switzerland\\
$^3$ Kiepenheuer-Institut f\"ur Sonnenphysik (KIS), D-79104 Freiburg i.~Br., Germany}

\begin{abstract}
The numerical computation of reliable and accurate Stokes profiles is of great relevance in solar physics.
In the synthesis process, many actors play a relevant role: among them the formal solver, the discrete atmospheric model, and the spectral line.
This paper tests the performances of different numerical schemes in the synthesis of polarized spectra for different spectral lines and atmospheric models.
The hierarchy between formal solvers is enforced, stressing the peculiarities of high-order and low-order formal solvers.
The density of grid points necessary for reaching a given accuracy requirement is quantitatively described for specific situations.
\end{abstract}


% %%%%%%%%%%%%%%%%%%%%%%%%%%%%%%%%%%%%%%%%
\section{Introduction}\label{sec:sec1}
% %%%%%%%%%%%%%%%%%%%%%%%%%%%%%%%%%%%%%%%%
%
Polarization is a property of light that encodes a wealth of information about the physical state of the environment from which the light emerges, making the radiative transfer of polarized light a topic of great interest in remote sensing. The usual differential form of the radiative transfer equation of partially polarized light reads
\begin{equation}
  \frac{\rm d}{{\rm d} s}\mathbf I(s) 
  = -\mathbf K(s)\mathbf I(s) + \boldsymbol{\epsilon}(s)\,.
\label{eq:RTE}
\end{equation}
It is a system of first-order coupled inhomogeneous ordinary differential equations, where $\mathbf I=(I,Q,U,V)^{T}$ is the Stokes vector, $\mathbf K$ is the $4\times4$ propagation matrix, and $\boldsymbol{\epsilon}$ is the emission vector. All of these quantities depend on the spatial coordinate $s$ measured along the ray under consideration. 
To simplify the notation, the frequency dependence of these
quantities is omitted.

%Analytical solutions of Equation~\eqref{eq:RTE} exist for a few simple atmospheric models only,
%Analytical solutions of Equation~\eqref{eq:RTE} are known for a few simple atmospheric models only \citep[e.g.,][]{lopez1999b}.
%hence the need to rely on numerical approximations.

Equation~\eqref{eq:RTE} seldom has
solutions that can be expressed in an analytical form, so it is common to seek approximate
solutions by means of numerical methods.
The so-called formal solution of Equation~\eqref{eq:RTE} consists of the numerical evaluation of the Stokes vector, given knowledge of the propagation matrix and the emission vector at a discrete set of spatial points and of the boundary conditions \citep{auer2003}. 
%It is well-known that such variations depend on different ingredients: among them the spectral line, the atmosphere model, the polarizing physical effects, and the spatial scale.

In the general nonlinear problem, the emission vector and the propagation matrix depend on the Stokes vector.
In this case, the formal solution is supplemented by the statistical equilibrium equations or by redistribution matrices for scattering processes. The whole problem is usually solved iteratively involving a large number of formal solutions per single iteration\footnote{For each discretized frequency and ray direction.} \citep{trujillo_bueno2003,landi_deglinnocenti+landolfi2004}.
%a well-performing formal solver is of particular importance for
Another issue of growing importance is the massive Stokes profiles synthesis on atmosphere models from 3D radiative-magnetohydrodynamic (R-MHD) simulations.
Moreover, inversion techniques require repetitive evaluation of Equation~\eqref{eq:RTE} \citep{bellot2000,deltoro_iniesta2016}.

These computationally demanding problems call for an efficient and accurate integration of Equation~\eqref{eq:RTE}, %, where many actors play a relevant role. 
%A fundamental ingredient is certainly the numerical scheme used to solve Equation~\eqref{eq:RTE}.
in which the crucial role of the formal solver is well recognized.
For this reason,
\citet{janett2017a,janett2017b} characterized formal solvers in terms of order of accuracy, stability, and computational cost.
Furthermore, \citet{janett2018a} analyzed the stiffness of the radiative transfer equation of polarized light,
identifying the instability issues %related to the propagation matrix
and outlining the structure of pragmatic numerical schemes.
%extended the same numerical analysis to high-order formal solvers.
%However, the quality of the calculated Stokes spectrum does not only depend on the numerical scheme.
In practice, the discrete atmospheric model also plays a fundamental role in the formal solution.
%In practice, many radiative transfer calculations are carried out considering standard semi-empirical atmospheric models. But also time instants of atmospheric models from 3D R-MHD simulations are of interest.
On the one hand, most of formal solvers achieve satisfying accuracy when considering finely sampled smooth model atmospheres.
%In this case, the large number of grid points needed to sample the atmosphere increases the total computational cost.
On the other hand, most of formal solvers notoriously fail on intermittent atmospheric models with coarse spatial grids.
%These models frequently show steep gradients, shock fronts, and peaks in temperature.
The debate on the optimal formal solution for polarized light is definitely not over.
%and this paper aims to contribute to the discussion, investigating different real-life Stokes profiles synthesis.

Section~\ref{sec:sec4} describes the usual contexts of numerical syntheses of Stokes profiles.
Section~\ref{sec:sec3} presents a selected pool of numerical schemes for the numerical integration of Equation~\eqref{eq:RTE}. %, explaining the criteria that lead to this choice.
Section~\ref{sec:sec2} focuses on selected atmospheric models, paying particular attention to their sampling in terms of density of spatial grid points.
Section~\ref{sec:sec5} exposes numerical tests for different cases. %specific situations.
%aiming at clarifying the usual numerical issues arising in the problem.
Finally, Section \ref{sec:sec6} provides remarks and conclusions.
\section{Physics}\label{sec:sec4}
Many practical applications of radiative transfer for polarized radiation in solar and stellar atmospheres concern the problem of line formation in the presence of a magnetic field \citep{landi_deglinnocenti+landolfi2004}.
There are various ingredients that 
%are important in the formation of a spectral line vary depending on the nature of the atomic transition of interest \citep{delacruz_rodriguez2016}. These ingredients
can modify the inputs of the formal solution, i.e., the propagation matrix and emission vector coefficients at the different grid points and their relative spacing.
This section provides a brief overview on the most important physical aspects in which the formal solutions generally differ.
\vspace*{0.5cm}
\subsection{Spectral line}\label{subsec:4.4}
Spectral lines originate from very different atomic transitions that are differently sensitive to atmospheric physical conditions \citep{deltoro_iniesta2016}.
Solar spectral lines are usually analyzed to determine the conditions of the atmospheric regions in which they are formed,
hence the distinction between photospheric, chromospheric, transition region, and coronal spectral lines.
In fact, a spectral line samples a range of geometrical heights, which usually increase from the line wings to the line center.
The relative contribution of an atmospheric layer to an observed quantity is generally quantified using contribution functions \citep{magain1986,grossmann-doerth+al1988},
while response functions provide information about how changes in the atmospheric parameters modify the emergent Stokes spectrum \citep{grossmann-doerth+al1988,ruiz_cobo1992,bellot_rubio+al1998}.

\begin{figure}
\centering
\includegraphics[width=.475\textwidth]{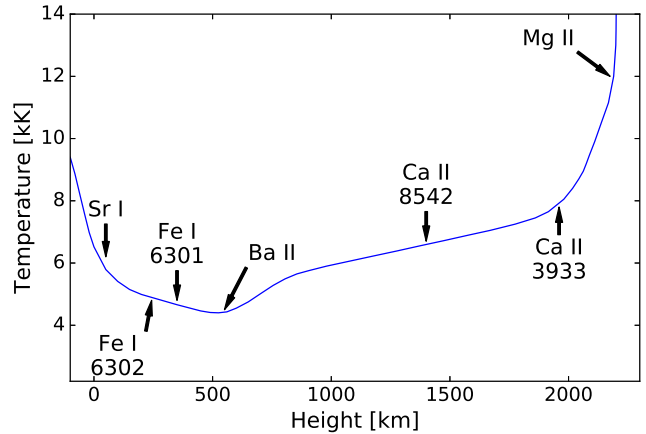}  
\caption{Formation heights for the spectral lines given in Table~\ref{tab:data} in the FALC atmospheric model.
Arrows indicate the height at which the line center optical depth is equal to unity.}
\label{fig:formation_heights}
\end{figure}

In this work, calculations are carried out for different spectral lines forming in photospheric and chromospheric regions. Table~\ref{tab:data} gives for each considered atomic transition the central wavelength, $\lambda_0$, and the Einstein coefficient, $A_{ul}$.
Figure~\ref{fig:formation_heights} shows the height
at which the line center optical depth is equal one for the selected spectral lines and atmospheric model. 
%(the subscript $\ell$ indicates that the optical depth is calculated at the line core wavelength).
%the multiplicity of the lower level times the oscillator strength, $\log (gf)$
%
\begin{table}
\caption{Spectral lines}
\setlength{\tabcolsep}{5pt}\renewcommand{\arraystretch}{1.5}
\centering
\begin{tabular}{|c|c|c|c|c|}\hline %\toprule
\emph{Ion}	& \emph{Transition} & $\lambda_0[{\rm \mbox{\AA}}]$	& $A_{ul}[s^{-1}]$	\\%$& $g_{eff}$	\log(gf)$
\hline %\midrule
Sr~{\sc i}	& $^1 S_0\, - \,^1P_1$		& 4607.3	& 2.01$\,\times\, 10^{8}$ 		\\%&?	\\%0.283		
Fe~{\sc i}	& $^5 P_2\, - \,^5 D_2$		& 6301.5	& 7.14$\,\times\, 10^{6}$		\\%&1.67	\\%-0.718		
Fe~{\sc i}	& $^5 P_1\, - \,^5 D_0$ 	& 6302.4	& 1.24$\,\times\, 10^{7}$					\\%&2.5	\\%-1.236
Ba~{\sc ii}	& $^2 S_{1/2}\, - \,^2 P_{3/2}$ & 4554.0	& 1.11$\,\times\, 10^{8}$		\\%&?	\\%0.140		
Ca~{\sc ii}	& $^2 D_{5/2}\, - \,^2 P_{3/2}$ & 8542.1	& 9.90$\,\times\, 10^{6}$		\\%&1.1	\\%-0.360
Ca~{\sc ii} K	& $^2 S_{1/2}\, - \,^2 P_{3/2}$ & 3933.6	& 1.47$\,\times\, 10^{8}$		\\%&?	\\%0.135		
Mg~{\sc ii} k	& $^2 S_{1/2}\, - \,^2 P_{3/2}$ & 2795.5	& 2.60$\,\times\, 10^{8}$		\\\hline%&?	\\		
%Ca~{\sc ii}	& $^2 D_{3/2}\, - \,^2 P_{1/2}$ & 8662.1	& 1.06e+07		&?	\\%-0.622 		
\end{tabular}
\vspace*{0.2cm}
\label{tab:data}
\end{table}
\subsection{LTE versus NLTE}\label{subsec:4.1}
The propagation matrix coefficients and the emission vector components in Equation~\eqref{eq:RTE}
are usually functions of the level population densities and other atmospheric physical parameters:
among them the temperature, the microscopic and macroscopic velocities, and the strength and orientation of the magnetic field. Under the assumption of local thermodynamic equilibrium (LTE), the level population densities are defined by the local temperature only and decoupled from the radiation field\footnote{In this case, only one integration of Equation \eqref{eq:RTE} is required to synthesize the emerging Stokes vector at each wavelength.}.
The assumption of LTE conditions satisfactorily describes the formation of many photospheric spectral lines. For lines forming higher up in the solar atmosphere, however, this assumption must be abandoned. This happens, for instance, in the presence of low collisional rates, where the atomic system and the radiation field interact in non-local thermodynamic equilibrium (NLTE) conditions. NLTE conditions imply that the emission vector and the propagation matrix depend in a complicated manner on the Stokes vector, so that Equation \eqref{eq:RTE} is nonlinear and the formal solution is supplemented by additional statistical equilibrium equations and/or by suitable redistribution matrices for scattering processes.

%In the formal solution for polarized light, the LTE or NLTE assumptions modify the propagation matrix and they could in principle alter the stiffness of the problem \citep[see][]{janett2018a}.
In this work, the LTE and NLTE conditions are assumed for the synthesis of photospheric and chromospheric lines, respectively.
NLTE effects are included in terms of the field-free approximation \citep{rees1969},
where the statistical equilibrium
equations for the level populations are solved as if the magnetic field was absent.
\subsection{Dimensionality}\label{subsec:4.2}
It is known that
``once the assumption of LTE is abandoned, the dimensionality of the atmosphere comes into play, since the non-locality of the radiation field manifests itself in the vertical,
as well as in the horizontal dimensions'' \citep{delacruz_rodriguez2017}.
\citet{mihalas+auer1978} and, subsequently, \citet{kunasz1988} presented the difficulties inherent
to the solution of unpolarized radiative transfer in 2D slab geometries.
\citet{stenholm1977} pioneered calculations of multidimensional transfer of the Stokes vector in magnetic fluxtubes.
More recently, \citet{stepan+trujillo_bueno2013} published PORTA, a code able to solve the non-equilibrium problem
of the generation and transfer of polarized radiation in 3D solar atmospheres.

For 1D problems, the grid points lay on the unique ray under consideration.
However, for 2D or 3D problems, the ray path (long characteristics) or the downwind and upwind points (short characteristics)
do not generally coincide with the grid points\footnote{Another option is the so-called 1.5D geometry,
in which each possible line of sight through the model atmosphere is treated independently, assuming for each a plane parallel atmosphere \citep[e.g.,][]{stenflo1994,pereira2015}.}.
% However, the ray path does not generally go through the original grid points for 2D or 3D problems.
In this case, one must recover the relevant quantities at off-grid points typically through interpolation (or reconstruction).
Such interpolations may introduce numerical errors in the absorption and emission coefficients.
Moreover, the cell widths of redefined discretized path 
may become highly irregular \citep[e.g.,][]{mihalas+auer1978}.

Since Equation~\eqref{eq:RTE} is always numerically integrated along a ray,
the dimensionality of the grid enters the formal solution only when %accessing the data only 
determining downwind and upwind quantities \citep{auer2003}.
Therefore, without any limitation to the analysis of formal solvers, the 1D geometry is always adopted in Section~\ref{sec:sec5}.
\subsection{Polarizing mechanisms}\label{subsec:4.3}
Polarization is introduced in spectral lines by several different symmetry-breaking mechanisms that are connected either to the presence of an external field,
e.g, the magnetic field, or to some kind of anisotropy in the excitation of the atomic system, e.g., the optical pumping \citep{landi_deglinnocenti+landolfi2004}.
The magnetic field influences the polarization of the emergent radiation in different ways, which depend on details of the atomic physics.
A common distinction is made between the Zeeman or Paschen-Back effects and the Hanle effect, which have different sensitivities to the strength and orientation of the magnetic field.
In the Stokes syntheses of this work, the polarization of spectral lines is produced by the Zeeman effect alone.
%Additional continuum linear polarization resulting from scattering by electrons (Thomson) and atoms (Rayleigh) is taken into account.
%
\subsection{Spatial scale}\label{subsec:4.5}
\begin{figure*}
\centering
\includegraphics[width=1.\textwidth]{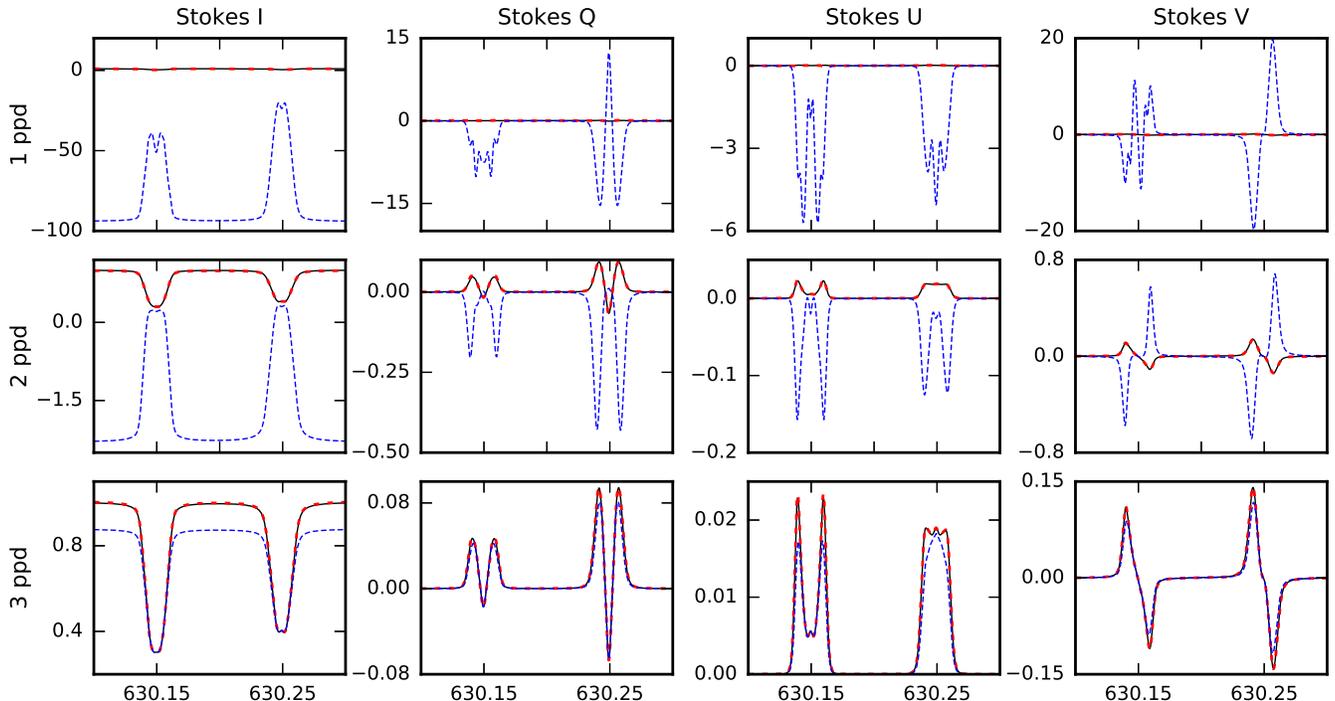}  \caption{Emerging Stokes vector components for the Fe~{\sc i} spectral lines at 6301.5 and 6302.5 {\rm \AA} in the FALC atmospheric model presented in Figure~\ref{fig:models}. The Stokes profiles are calculated with the cubic Hermitian method in the geometrical depth scale (dashed blue) and in the optical depth scale (dashed red) for different spatial point densities.
The reference emergent Stokes profile (continuous black) is calculated using a sampling with $10^3$ points-per-decade (ppd) of continuum optical depth.
The dashed red and solid black lines almost overlap. Note the variations in the $y$-axis scale among the different plots.}
\label{fig:conversion}
\end{figure*}

%
%Recently, \citet{janett2018a} investigated the stiffness of the polarized radiative transfer equation and the inherent instability encountered by numerical schemes.
\citet{janett2018a} demonstrated that the stiffness of Equation~\eqref{eq:RTE} is spatial scale dependent.
The conversion to the optical depth scale given by
\begin{equation}
\tau(s) = \tau_0 + \int_{s_0}^s \eta_I(x)\,{\rm d}x\,,
\label{conversion_opt_depth}
\end{equation}
where $\eta_I$ is the intensity absorption coefficient at the frequency under consideration,
recasts Equation~\eqref{eq:RTE} into the equivalent form\footnote{The sign convention is according to \citet{janett2018a}.}
\begin{equation*}
\frac{\rm d}{\rm d\tau}\mathbf I(\tau) = -\tilde{\mathbf K}(\tau)\mathbf I(\tau) + \tilde{\boldsymbol{\epsilon}}\,,
\end{equation*}
where $\tilde{\mathbf K}=\mathbf K/\eta_I$ and $\tilde{\boldsymbol{\epsilon}}=\boldsymbol{\epsilon}/\eta_I$.
The use of the optical depth scale %defined by Equation~\eqref{conversion_opt_depth}
(instead of the geometrical scale)
usually mitigates variations of the entries of the propagation matrix along the ray path, damping instability issues.
Figure~\ref{fig:conversion} presents some numerical evidence,
showing the unstable (stable) behavior of the cubic Hermitian method when using the geometrical (optical) depth scale on very coarse spatial grids\footnote{
This fact explains the contradicting opinions on the numerical stability of the cubic Hermitian method present in the literature
\citep[e.g.,][]{bellot_rubio+al1998,piskunov2002,delacruz_rodriguez+piskunov2013}.}.
Note that the conversion to the optical depth scale requires the quadrature of the integral in Equation~\eqref{conversion_opt_depth}
and introduces numerical errors. These errors might reduce the order of accuracy of the formal solver
and even lead to instabilities issues. 
%In order to maintain high-order convergence, high-order methods require a corresponding high-order spatial scale conversion.
% 
% \citet{landi_deglinnocenti1976} propose as an alternative to perform the conversion to the $\log(\tau)$ scale, pointing out that the atmospheric parameters vary more smoothly in this spatial scale. However, a more in-depth study is required to support this proposition.

For the sake of clarity, calculations in Section~\ref{sec:sec5} are carried out in the optical depth scale given by Equation~\eqref{conversion_opt_depth} at the frequency under consideration (with the exception of the pragmatic formal solvers).
% % 
High-order numerical schemes are always combined with corresponding order monotonic spatial scale conversion: trapezoidal rule for second-order schemes and
interpolatory quadrature rules for third-order \citep[based on][]{fritsch1984} and forth-order \citep[based on][]{steffen1990} methods.
\section{Formal solvers}\label{sec:sec3}
% %%%%%%%%%%%%%%%%%%%%%%%%%%%%%%%%%%%%%%%%
%An extensive literature on formal solvers for the polarized radiative transfer exists.
For the sake of clarity, only a limited number of numerical schemes is analyzed.
However, the numerical tests performed in Section~\ref{sec:sec5} can be unconditionally applied to any formal solver.
This section presents a pool of candidate numerical schemes selected according to three criteria: order of accuracy, numerical stability, and computational cost.
Table~\ref{tab:convergence} gives a short overview of such schemes\footnote{
The computational cost depends on the specific coding, programming language, compiler, and computer architecture. Nevertheless, the computational time per step given in Table~\ref{tab:convergence} is a good indicator for the complexity of the numerical scheme.}. 
\begin{table}
\caption{Numerical schemes}
\setlength{\tabcolsep}{5pt}\renewcommand{\arraystretch}{1.5}
\centering
\begin{tabular}{|l|c|c|c|}\hline%\toprule
\emph{Formal solver}	& \emph{Order}	& \emph{Stability}	& \emph{Time cost}\footnote{Second per step averaged over $2\times 10^{6}$ steps.}\\
\hline %\midrule
Heun's 			& 2 	& -		&2.51$\times 10^{-7}$\\
Trapezoidal 		& 2 	& $A$-stable	&8.40$\times 10^{-7}$\\
DELO-linear		& 2 	& $L$-stable\footnote{\label{note1}$L$-stability is guaranteed for diagonal $\mathbf K$.}	&9.65$\times 10^{-7}$\\
Runge-Kutta 3		& 3 	& - 		&3.83$\times 10^{-7}$\\
Adams-Moulton 3		& 3 	& - 		&1.05$\times 10^{-6}$\\
DELO-parabolic		& 3 	& $L$-stable\textsuperscript{\ref{note1}} 	&1.15$\times 10^{-6}$\\
cubic Hermitian		& 4 	& $A$-stable 	&1.15$\times 10^{-6}$\\
%Runge-Kutta 4		& 4 \\
quadratic DELO-B\'ezier	& 4 	& $L$-stable\textsuperscript{\ref{note1}}	&1.35$\times 10^{-6}$\\
cubic DELO-B\'ezier	& 4 	& $L$-stable\textsuperscript{\ref{note1}}	&1.40$\times 10^{-6}$\\\hline 
\end{tabular}
\vspace*{0.3cm}
\label{tab:convergence}
\end{table}%\vspace{0.1cm}
\subsection{Trapezoidal method}\label{subsec:3.1}
The well-known implicit trapezoidal method is an optimal representative of low-order implicit formal solvers.
It is second-order accurate and $A$-stable (i.e., its stability region comprises the complex left half-plane $\mathbb{C}^{-}$).
Moreover, it consists of a skeletal algorithm.
%and the lack of high-order accuracy can possibly be balanced by low computational cost.
\citet{janett2017a} gave a detailed description of the method and its application to Equation~\eqref{eq:RTE}.
%, where a direct comparison to DELO-methods is also provided.
%
\subsection{Adams-Moulton 3 method}\label{subsec:3.2}
The Adams-Moulton 3 method is an implicit linear two-step method and it is third-order accurate.
It suffers from numerical instability when treating optically thick cells because of its bounded stability region \citep{janett2017b}.
Moreover, the effect of variable cell size on the stability of multistep methods is not fully understood
and might be relevant for radiative transfer applications.
A deeper look into this problem is given by \citet{gear1974} and \citet{grigorieff1983}.
%, where a direct comparison to different high-order methods is also provided.
%
\subsection{Cubic Hermitian method}\label{subsec:3.5}
\citet{bellot_rubio+al1998} first proposed the cubic Hermitian method for the integration of Equation~\eqref{eq:RTE}.
%, highlighting its high performance.
\citet{janett2017b} subsequently reaffirmed the excellence of this implicit method in terms of accuracy and stability: it is fourth-order accurate and $A$-stable.
A possible weakness of this  method could lay in its complexity. In fact, it requires matrix-by-matrix multiplications
and, in particular, the calculation of second-order accurate numerical derivatives appears to be the most expensive part of the method.

Moreover, \citet{ibgui2013} explained that the calculation of numerical derivatives proposed by
\citet{steffen1990} is sensibly slower than the one proposed by \citet{fritsch1984}.
However, the derivatives provided by \citet{fritsch1984} are second-order accurate on uniform
grids only and they drop to first-order accuracy on non-uniform grids \citep[see][]{janett2018a}.
For this reason, the version by \citet{steffen1990} is used here.
% \citet{janett2017b} showed that second-order accurate numerical derivatives for K and
%   \epsilon are necessary to rach fourth-order accuracy with the cubic Hermitian method.}
%
% \subsection{RK4 method}\label{subsec:3.6}
% %
% The fourth-order accurate RK4 method is here applied in terms of the hybrid technique proposed by \citet{janett2017b}: it switches to the A-stable trapezoidal method when dealing with optically thick cells, trying to increase stability, maintaining high-order convergence. The RK4 method is an explicit numerical scheme, avoiding the solution of the $4\times4$ linear system and reducing the computational cost. A deeper description of the method, including its application to Equation~\eqref{eq:RTE}, can be found in \citet{janett2017b}, where a direct comparison to different high-order methods is also provided.
% 
%
\subsection{DELO-linear}\label{subsec:3.2b}
The popular DELO-linear implicit formal solver
linearly approximates the effective source function \citep[see][]{rees+al1989} guaranteeing second-order accuracy.
Thanks to its proximity to $L$-stability, it avoids instability issues and
%Like any other method belonging to the DELO family, it uses exponential functions, which burden the algorithm.
correctly replicates exponential attenuations even when the step-size is large \citep{janett2018a}.
\subsection{DELO-parabolic}\label{subsec:3.3}
The DELO-parabolic implicit method\footnote{This method should not be interchanged with the DELOPAR method by \citet{trujillo_bueno2003},
which is only second-order accurate.} approximates the effective source function with a parabola,
resulting in a third-order accurate two-step method \citep{janett2017a}.
%The higher-accuracy with respect to the DELO-linear method is however counterbalanced by a diminished stability.
There exist contradicting opinions on its stability properties:
\citet{murphy1990} mentioned the numerical instability due to the parabolic approximation (overshooting),
while \citet{janett2018a} demonstrated its proximity to $L$-stability.
Similarly as for the Adams-Moulton~3 method, the effect of variable cell size might be relevant for stability issues.
%and a more detailed description of this formal solver is given in \citet{janett2017a}.
%
\subsection{Quadratic and cubic DELO-B\'ezier methods}\label{subsec:3.4}
The quadratic and cubic DELO-B\'ezier  methods are implicit schemes based on quadratic and cubic B\'ezier approximations of the effective source function, respectively \citep{delacruz_rodriguez+piskunov2013}.
Providing at least second-order accurate derivatives, both methods can reach fourth-order accuracy.
\citet{janett2018a} demonstrated their proximity to $L$-stability.
Similarly to the cubic Hermitian method, DELO-B\'ezier methods require the calculation of numerical derivatives
that increases the total computational effort.
For the same arguments given in Section~\ref{subsec:3.5}, the calculation of numerical derivatives proposed by
\citet{steffen1990} is preferred here.
%Moreover,  and \citet{janett2017a} gave a detailed description of the method and its application to Equation~\eqref{eq:RTE}.
% 
%
\subsection{Second-order pragmatic method}
\citet{janett2018a} suggested considering each integration interval at a time and sequentially.
% In each interval, depending on the cell width $\Delta s$ and on the magnitude of the
% eigenvalues of $\mathbf{K}$ at $s_i$ and $s_{i+1}$,
In each interval, depending on the cell width $\Delta s$ and on the
eigenvalues of $\mathbf{K}$ at the cell boundaries,
the Stokes vector is calculated with either
an explicit method $\mathbf{\Psi}^E$ (which is computationally cheaper) % inexpensive),
or an $A$-stable method $\mathbf{\Psi}^A$, or an $L$-stable method $\mathbf{\Psi}^L$.

The second-order pragmatic method proposed here uses
Heun's method as $\mathbf{\Psi}^E$, the implicit trapezoidal method as $\mathbf{\Psi}^A$,
and the DELO-linear method  as $\mathbf{\Psi}^L$.
Both Heun's method and the implicit trapezoidal method use the geometrical spatial scale. 
The method $\mathbf{\Psi}^E$ is used if the stability function $\phi_{\mathbf{\Psi}^E}$ satisfies $\phi_{\mathbf{\Psi}^E}<0.95$
or if $\Delta\tau<10^{-3}$;
alternatively, the method $\mathbf{\Psi}^A$ is used if $\phi_{\mathbf{\Psi}^A}<1.0$ \citep[see][Sect. 6]{janett2018a}.
Otherwise, or whenever $\Delta\tau>7$, the method $\mathbf{\Psi}^L$ is used,
guaranteeing numerical stability and the correct exponential attenuation of the Stokes vector.
%The method $\mathbf{\Psi}^A$ is converted to optical depth if $S_{\mathbf{\Psi}^A}<0.8$.
These parameters should not be considered as an ultimate choice,
but they provide accurate and stable numerical results.
Algorithm~\ref{alg1} in Appendix~\ref{appendix:B} presents the pseudocode of the method.

The overhead caused by the first switching decision is roughly half as expensive as one step of Heun's method.
Moreover, the use in percentage of $\mathbf{\Psi}^E$, $\mathbf{\Psi}^A$, and $\mathbf{\Psi}^L$
depends on the atmospheric model and on the density of grid points \citep[see][Table 2]{janett2018a}.
For these reasons, the pragmatic method is particularly competitive when it performs many steps with $\mathbf{\Psi}^E$,
but it looses efficiency when using many $\mathbf{\Psi}^A$ and $\mathbf{\Psi}^L$ steps, i.e., in very coarse grids.
\subsection{Third-order pragmatic method}
%
%This is the third-order version of the method proposed above.
The pragmatic strategy by \citet{janett2018a} is extensible to higher order.
For instance, the third-order pragmatic method proposed here uses the explicit
RK 3 method as $\mathbf{\Psi}^E$ \citep[e.g.,][]{hairer2000},
the third-order Hermitian method\footnote{It corresponds
to the fourth-order cubic Hermitian method by \citet{bellot_rubio+al1998} with first-order numerical derivatives.} as $\mathbf{\Psi}^A$,
and the DELO-linear method  as $\mathbf{\Psi}^L$\footnote{The method
$\mathbf{\Psi}^L$ can be of lower order because it is applied in optically thick regions, which usually prevent any propagation of information.}.
Both the RK 3 method and the third-order Hermitian method use the geometrical spatial scale.
The method $\mathbf{\Psi}^E$ is used if the stability function $\phi_{\mathbf{\Psi}^E}$ \citep{frank2008} satisfies $\phi_{\mathbf{\Psi}^E}<0.95$
or if $\Delta\tau<10^{-3}$;
alternatively, the method $\mathbf{\Psi}^A$ is used if $\phi_{\mathbf{\Psi}^A}<1.0$ \citep[see][]{janett2017b}.
Otherwise, or whenever $\Delta\tau>10$, the method $\mathbf{\Psi}^L$ is used,
guaranteeing numerical stability and the correct exponential attenuation of the Stokes vector.
Algorithm~\ref{alg2} in Appendix~\ref{appendix:B} presents the pseudocode of the method.

The computational efficiency of this method is analogous to that of the second-order pragmatic method.
% %%%%%%%%%%%%%%%%%%%%%%%%%%%%%%%%%%%%%%%%
\section{Atmospheric models}\label{sec:sec2}
%
%Depth-dependent properties of the solar atmosphere are generally studied through the analysis of spectral lines and the different atmospheric regions have different importance for the creation of Stokes signals.
Radiative transfer problems are usually solved for discrete atmospheric models,
where the physical quantities describing the atmosphere are given at a discrete set of spatial points.
The propagation matrix and emission vector coefficients are then calculated at grid points, allowing the numerical integration of Equation~\eqref{eq:RTE}.

Equation~\eqref{eq:RTE} is usually solved on the atmospheric model grid.
However, one can still add, move, or remove grid points: this is usually done through interpolation and by cutting off irrelevant atmospheric layers.
%One usually aims at using spatial grids as coarse as possible, maintaining the numerical error under a certain limit value.
It is certainly not possible to state the minimal amount of grid points necessary to achieve a certain accuracy
because an atmosphere sampling giving satisfactory results for one specific spectral line may produce large errors for another one formed at a different depth and range.
% In practice, only some atmospheric layers effectively contribute to the formation of a specific spectral line, which otherwise is fairly sensitive to other atmospheric layers.
Clearly, it is self-defeating to distribute spatial grid points too deep in the solar atmosphere, 
where the exponential attenuation cancels any contribution to the emergent spectrum. 
% In fact, the usual attenuation operator depends exponentially on the optical depth. 
When considering the homogeneous version of Equation~\eqref{eq:RTE} in the limit of a vanishing magnetic field, one has
\begin{equation*}
\mathbf I(\tau)\approx e^{-\Delta\tau}\mathbf I(\tau_0)\,,
\label{exp_att}
\end{equation*}
where $\Delta\tau=|\tau-\tau_0|$. This exponential attenuation guarantees that an optically thick layer, say, $\Delta\tau\ge20$,
cancels any relevant contribution of the incoming light to the emergent spectrum. 
A quantitative perception of the exponential attenuation is given in Table~\ref{tab:attenuation}.
%Moreover, layers with vanishing emission and absorption can be omitted.

%In order to compute the global error of the formal solution as a function of the ppd,
%the atmospheric models are here re-sampled with different spatial point densities. 
When analyzing the performances of formal solvers, it is common to choose a homogeneous sampling in $\log\tau_c$ \citep[e.g.,][]{rees+al1989,bellot_rubio+al1998,delacruz_rodriguez+piskunov2013}, where the coordinate $\tau_c$ is the continuum optical depth defined by
\begin{equation*}
{\rm d}\tau_c=-\eta_c(s){\rm d}s\,,
\label{opt_depth}
\end{equation*}
with $\eta_c$ being the continuum absorption coefficient at the wavelength $\lambda=5000$ {\rm \AA}.
One then defines the quantity of points-per-decade (ppd) of $\tau_c$, indicating the amount of grid points that sample a variation of one order of magnitude in the continuum optical depth.
Therefore, one speaks of the grid-point density (ppd) rather  than referring to the total amount of grid points.

In order to analyze the accuracy of the formal solution as a function of the cell width,
one needs a sequence of discrete grids with an increasing (or decreasing) grid-point density.
In this work, the following procedure is used:
%and a homogeneous sampling in $\log\tau_c$.
first, the original discrete atmospheric model is interpolated in terms of cubic splines in each quantity\footnote{
Cubic splines guarantee that variation in the atmospheric models are smooth enough, such that high-order methods can actually reach high-order convergence.}.
Second, the run of the atmospheric parameters is restricted to a specific interval in continuum optical depth (e.g., $\log\tau_c\in[-8,2]$).
Third, a sequence of models with increasing grid-point densities is generated. 

In the following, three different atmospheric model categories are presented: analytical, semi-empirical, and 3D R-MHD atmospheric models.
\begin{table}
\caption{Exponential attenuation}
\setlength{\tabcolsep}{5pt}\renewcommand{\arraystretch}{1.5}
\centering
\begin{tabular}{|c|c|c|}\hline%\toprule
Optical width $\Delta \tau$	& $\log \Delta \tau$	& Attenuation factor $e^{-\Delta \tau}$\\
\hline %\midrule
1	& 0.0	& 0.36 \\
5	& 0.7	& 6.7$\times 10^{-3}$ \\
10	& 1.0	& 4.5$\times 10^{-5}$ \\
20	& 1.3	& 2.0$\times 10^{-9}$ \\
50	& 1.7	& 1.9$\times 10^{-22}$ \\
100	& 2.0	& 3.7$\times 10^{-44}$\\\hline
\end{tabular}
\label{tab:attenuation}
\end{table}\vspace*{0.2cm}
\begin{figure*}
\centering
\includegraphics[width=1.\textwidth]{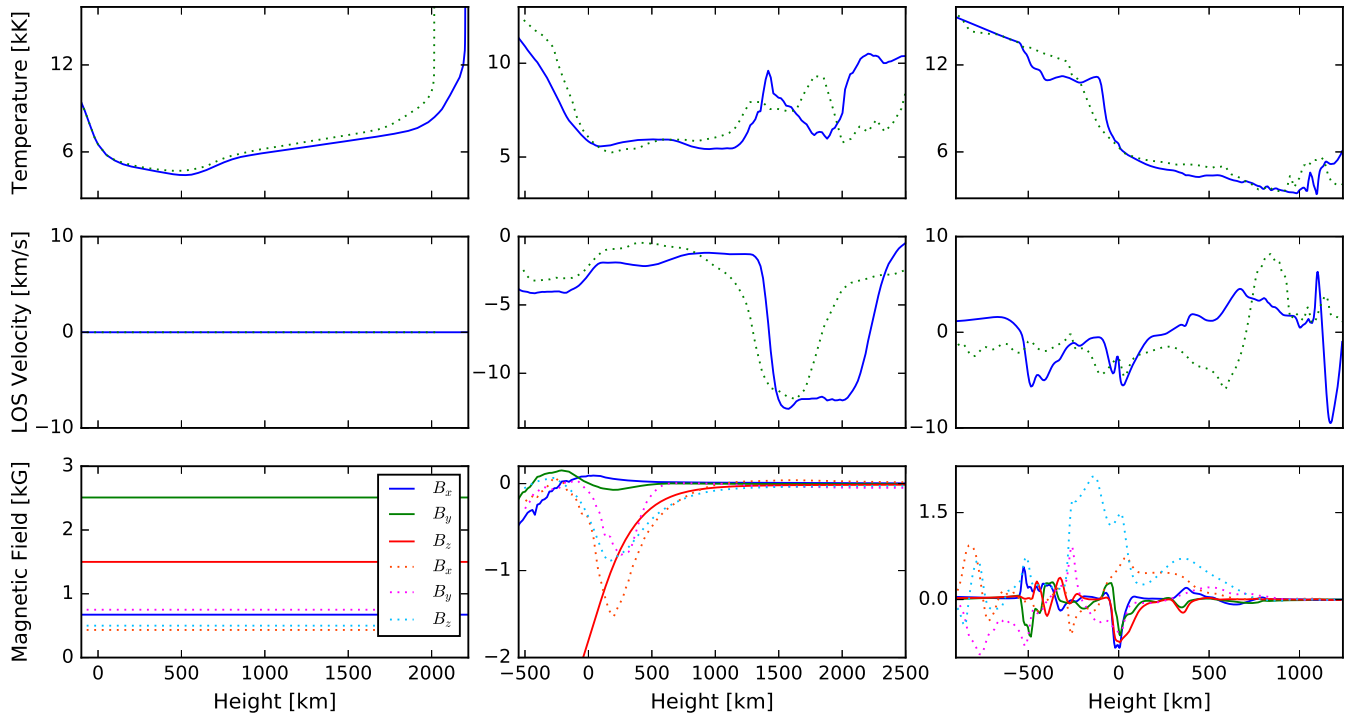}
% \caption{The upper row shows the temperature stratification as a function of height. The middle row shows the macroscopic l.o.s velocity. The bottom row illustrates the three components of the magnetic field. Considered atmospheric models: first column FALC (continuous), FALF (dotted);  second column Bifrost-1 (continuous), Bifrost-2 (dotted); third column  CO$^5$BOLD-1 (continuous),  CO$^5$BOLD-2 (dotted).}
\caption{Temperature (top row), upward directed line of sight velocity (middle row), three components of the magnetic field (bottom row), all as a function of the geometrical height in the atmosphere.
Considered atmospheric models: first column FALC (solid), FALF (dotted);  second column Bifrost-1 (solid), Bifrost-2 (dotted); third column  CO$^5$BOLD-1 (solid),  CO$^5$BOLD-2 (dotted).
Note that Bifrost-1, Bifrost-2, CO$^5$BOLD-1, and CO$^5$BOLD-2 are single columns of time instants of 3D R-MHD simulations.}
\label{fig:models}
\end{figure*}
\subsection{Analytical atmospheric models}\label{subsec:2.2}
A great number of attempts have been made to obtain exact solutions of the polarized radiative transfer equation using analytical atmospheric models.
The best-known case is the so-called Milne-Eddington atmospheric model. This model implies LTE conditions, a constant propagation matrix,
and that the Planck function varies linearly with optical depth.
\citet{unno1956} first provided the solution of Equation~\eqref{eq:RTE} for a Milne-Eddington atmosphere with vanishing magneto-optical effects, and
\citet{rachkovsky1967} subsequently gave the complete solution.
% Despite of its simple schematization of a real stellar atmosphere, the Milne-Eddington atmospheric model is still used, e.g., for inferring average values of the magnetic field and the l.o.s velocity \citep{deltoro_iniesta2016}.
Thereafter, \citet{landi_deglinnocenti1987} analytically solved Equation~\eqref{eq:RTE} in terms of the evolution operator for different cases, where both the propagation matrix and the emission vector are given functions of the optical depth. In addition, \citet{lopez1999b} propose an alternative strategy to analytically handle different cases with a non-constant propagation matrix.
Moreover, analytical atmospheric models have been frequently used to test the performances of various formal solvers \citep[e.g.,][]{bellot_rubio+al1998,stepan+trujillo_bueno2013}. In particular, \citet{janett2017a,janett2017b} tested different formal solvers using an analytically prescribed model atmosphere.
To avoid superfluous repetitions, no analytical atmospheric model is considered in this work.
\subsection{Semi-empirical atmospheric models}\label{subsec:2.3}
Semi-empirical models  %\citep[e.g.,][]{belluzzi2015,alsinaballester2017}.
% describe the variations of the essential parameters across the atmosphere, as a function of the height (or column mass)
% and are frequently used to synthesize solar spectral lines.
% These models
assume a 1D atmosphere in hydrostatic equilibrium with a temperature structure,
% and provide a mean state of 
so as to reproduce various mean diagnostic quantities of
the atmosphere, averaging both temporally and spatially \citep{mauas2007}.

There exist different solar atmospheric discrete models for the photosphere, for the chromosphere, for plage regions, for sunspot umbrae, etc.
The quiet-Sun atmospheric models of \citet{holweger1974} and the Harvard-Smithsonian reference atmospheric model of \citet{gingerich1971} together with their updated versions of \citet{vernazza1981} and \citet{fontenla1990,fontenla1991,fontenla1993,fontenla1999}
are some of the best-known semi-empirical atmospheric models. The latter models, better known by the acronym FAL, 
%contain around 70 grid points,
sample the solar atmosphere from the photosphere up to the transition region.
%, in the interval $\log\tau_c\in[-10,2]$.

The first column of Figure~\ref{fig:models} shows the variation of different atmospheric physical quantities along the vertical for FALC and FALF.
FAL models do not include a specific magnetic field. For this reason, a constant magnetic field is considered\footnote{The reason for the large magnetic field is the requirement of sufficiently large $Q$, $U$, and $V$ Stokes components to limit the undesirable loss of significance due to finite-precision in numerical tests.}.
Figures~\ref{fig:ppd}a (quiet-Sun FALC model) and~\ref{fig:ppd}b (network FALF model) clearly show that the atmosphere
is not homogeneously sampled in $\log \tau_c$ with strong variations in the grid-point density. For instance, the interval $\log\tau_c\in[-3,2]$
is usually sampled with only 2-4 ppd, while in the upper layers the sampling reaches 5-14 ppd.
This sampling reflects the availability and abundance of diagnostics from which the semi-empirical atmosphere is constructed. 
%
%The numerical results obtained with standard non-homogeneous samplings (see Section~\ref{sec:sec2}
%and Figures~\ref{fig:fal_ppd} and \ref{fig:mhd_ppd}) are presented in Appendix~\ref{appendix:A}.
%
\subsection{R-MHD atmospheric models}\label{subsec:2.4}
State-of-the-art R-MHD simulations are expected to realistically represent the solar atmosphere and allow for full 3D polarized radiative transfer calculations.
However, the sampling of stellar atmospheric models from 3D R-MHD simulations is determined by the needs of the numerical evaluation of the MHD systems of partial differential equations and it may not be optimal for the synthesis of Stokes profiles. Here, two different examples of 3D R-MHD atmospheric models are considered.
These are two single columns, 1 and 2, of time instants of Bifrost and CO$^5$BOLD simulations, respectively.

\begin{figure*}
\centering
\includegraphics[width=1.\textwidth]{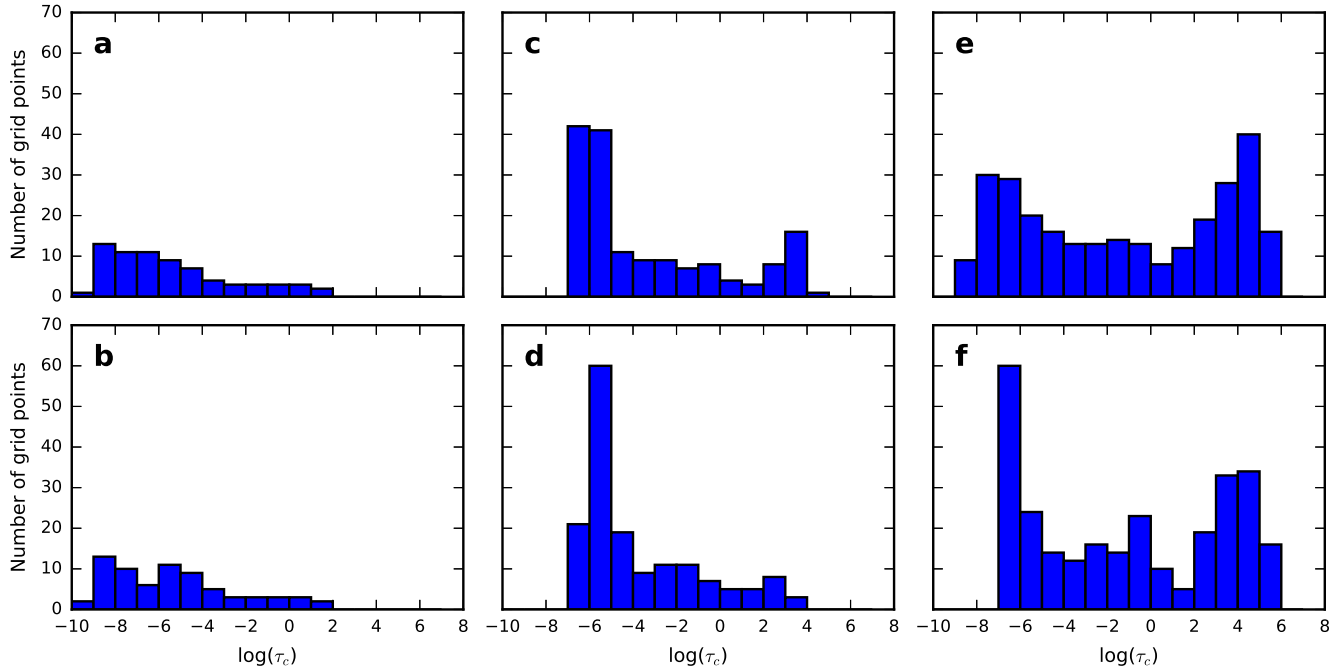}  \caption{Histograms of grid-point density for the different atmospheric models considered here: {\bf a)} FALC, {\bf b)} FALF, {\bf c)} Bifrost-1, {\bf d)} Bifrost-2, {\bf e)} CO$^5$BOLD-1, and {\bf f)} CO$^5$BOLD-2.}
\label{fig:ppd}
\end{figure*}

Bifrost 3D R-MHD models of the solar atmosphere including non-equilibrium hydrogen ionization \citep[see e.g.,][]{gudiksen2011,carlsson2016} are frequently used when investigating chromospheric phenomena \citep[e.g.,][]{stepan2015,carlin2017}. The second column of Figure~\ref{fig:models} shows the variation of different atmospheric physical quantities
for two vertical columns of a Bifrost snapshot\footnote{Model $en024048\_hion$.}. 
Figures~\ref{fig:ppd}c and~\ref{fig:ppd}d show that the atmosphere is not homogeneously sampled in $\log \tau_c$ and strong variations
in the grid-point densities appear. The interval $\log\tau_c\in[0,2]$ is sampled with only 3-5 ppd, reaching up to 60 ppd in other layers. 
%A finer sampling of more than 30 ppd is present in the interval $\log\tau_c\in[-7,-5]$.

CO$^5$BOLD 3D R-MHD simulations are mainly intended to describe the photospheric regions \citep[see e.g.,][]{freytag2012,calvo2016}. The third column of Figure~\ref{fig:models} shows the variation of different atmospheric physical quantities for two vertical columns of a CO$^5$BOLD snapshot\footnote{Model $d3f57g45h50fcn116$ of \citet[][Tab. 3]{steiner2017}, columns $(ix,iy)=(518,669)$ and $(233,209)$.}.
In this simulation, horizontal magnetic field was advected into the computational domain resulting in particularly turbulent and intermittent magnetic fields.
Figures~\ref{fig:ppd}e and~\ref{fig:ppd}f show that the atmosphere is non-homogeneously sampled in $\log \tau_c$, with strong variations in the grid-point density.
The interval $\log\tau_c\in[1,2]$ is sampled with less than 10 ppd.
In the other layers the sampling lays between 10-15 ppd,
with the exception of a finer sampling of more than 25 ppd in the intervals $\log\tau_c\in[-7,-5]$ and $\log\tau_c\in[3,5]$.
% 
% In both Bifrost and CO$^5$BOLD atmospheric models, an important amount of spatial points is distributed deep in the atmosphere, where the contribution to the emergent Stokes profiles is irrelevant. These layers can be simply omitted, lightening the computational problem.
%
%%%%%%%%%%%%%%%%%%%%%%%%%%%%%%%%%%%%%%%%
\section{Numerical results}\label{sec:sec5}
The Stokes-profile syntheses are performed using a modified version of the RH code of \citet{uitenbroek2001}
based on the MALI formalism of \citet{rybicki1991,rybicki1992,rybicki1994}.
The code, written in the C language, solves the combined equations of statistical equilibrium and radiative transfer
for multilevel atoms and molecules in a given plasma under general NLTE conditions. 
The modified version allows to switch between the different formal solvers presented in Section~\ref{sec:sec3} and to sequentially perform Stokes-profile syntheses 
with a set of discrete atmospheric models.
In these calculations, the polarization of spectral lines is produced by the Zeeman effect alone.

% For spectral lines forming in NLTE conditions, the level population densities are iteratively computed for the (hyperfine)
% reference atmospheric models in terms of the field-free approximation.
% Re-samplings of each level population density are then performed for the sequences of atmospheric models with different grid-point density.
NLTE effects are taken into account in terms of the field-free approximation,
which iteratively computes the level population densities for the unpolarized case (ignoring the magnetic field) with a third-order formal solver. 
NLTE level population densities are affected by numerical errors that are generally higher for coarser grids.
In order to discount for such errors,
the level population densities are iteratively computed for the (hyperfine) reference atmospheric models only.
Re-samplings of each level population are then performed through interpolation for the sequences of atmospheric models with different grid-point densities.
The formal solvers presented in Section~\ref{sec:sec3} are then used to 
synthesize the emerging Stokes profiles with one single integration of Equation~\eqref{eq:RTE}.

The Stokes syntheses are performed under the approximation of complete redistribution (CRD) even for Ca~{\sc ii} K and the Mg~{\sc ii} k spectral lines.
This crude simplification is done in order to detect the error that originates from the formal solver alone and not also from iterative calculations 
of partial frequency redistribution (PRD).
% Subsequently, a re-sampling for each level population density with different grid-point densities (the same as described in Section~\ref{sec:sec2} for atmospheric models)
% is performed.
%The Rybicki \& Hummer formalism allows radiative (bound-free as well as bound-bound) transitions to overlap in wavelength. %\textcolor{orange}{The calculations are carried out on a laptop with: Memory 7.5 GiB. OS type 64-bit. Processor Intel Core i7-5500U CPU @ 2.40GHz $\times$ 4.}
%

The formal solvers presented in Section~\ref{sec:sec3} are tested\footnote{
Numerical tests show that the performances achieved by quadratic and cubic DELO-B\'ezier methods are almost identical.
For simplicity, only the results achieved by the cubic DELO-B\'ezier method are presented.} on the atmospheric models shown
in Figure~\ref{fig:models} with the spectral lines given in Table~\ref{tab:data}.
The atmospheric models are re-sampled homogeneously in $\log\tau_c$ with different grid-point densities (see Section~\ref{sec:sec2}):
the FAL atmospheric models are re-sampled in the optical depth interval $-8.6\le\log\tau_c\le1.4$, % which encompasses 10 decades,
the Bifrost vertical columns in $-6.5\le\log\tau_c\le 3.5$,
and the CO$^5$BOLD vertical columns in $-8\le\log\tau_c\le 2$. %which encompasses 8 decades.
Each spectral line is sampled with around 500 points equispaced in frequency
in a spectral interval of a few {\rm \AA} around the core.  The global error is defined by Equation~\eqref{error}.

\begin{figure*}
\hspace*{-0.5cm}
\includegraphics[width=.35\textwidth]{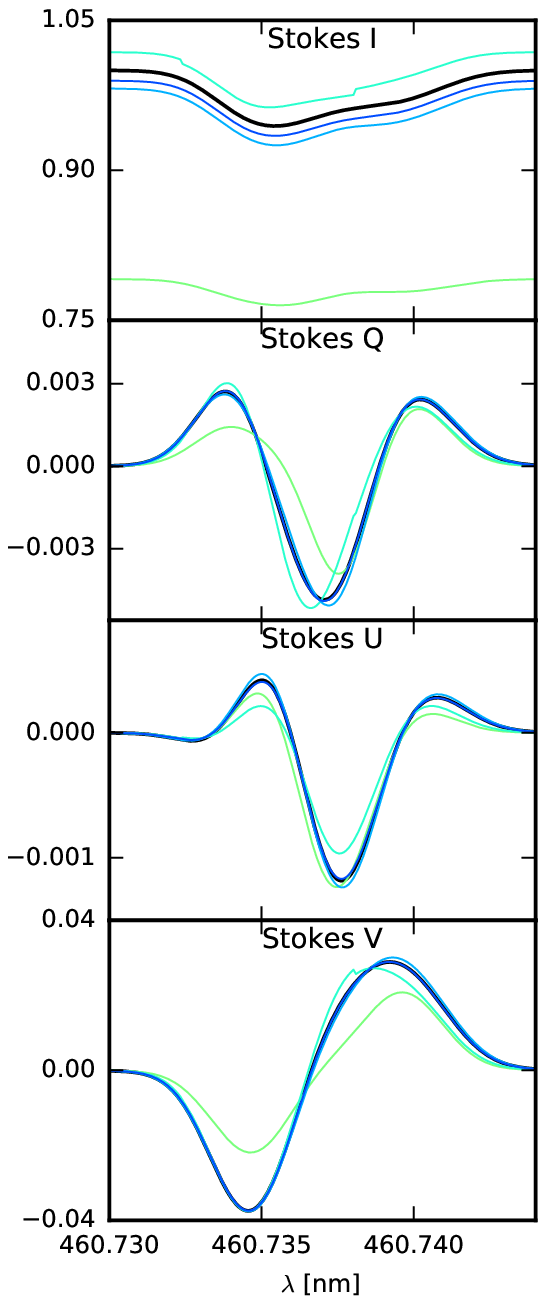}
\hspace*{-0.40cm}
\includegraphics[width=.35\textwidth]{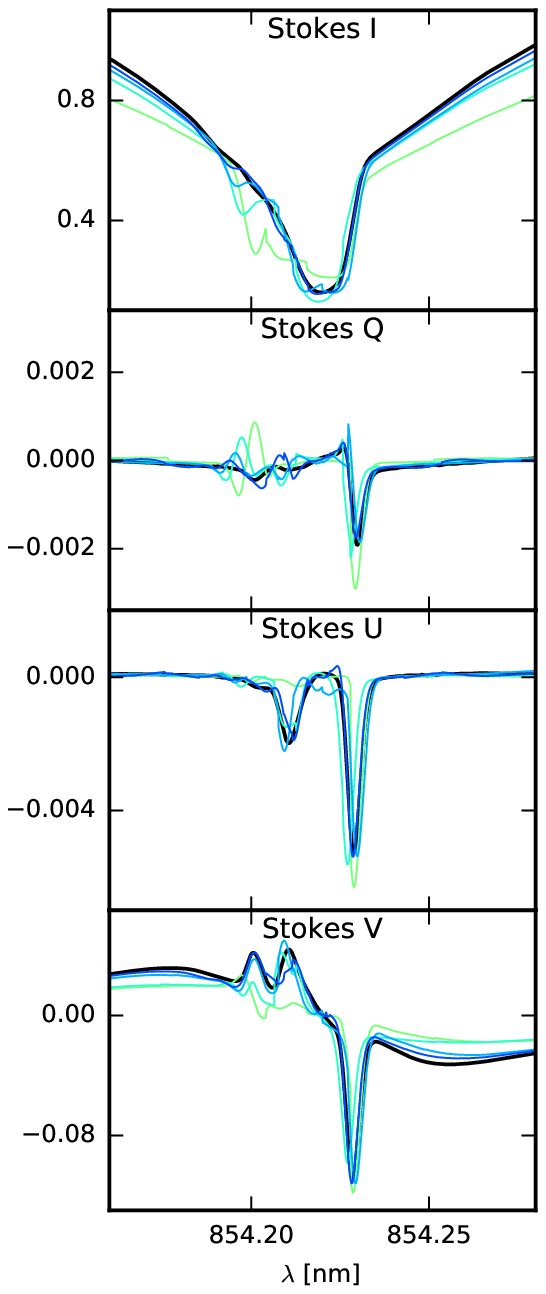}
\hspace*{-0.40cm}
\includegraphics[width=.35\textwidth]{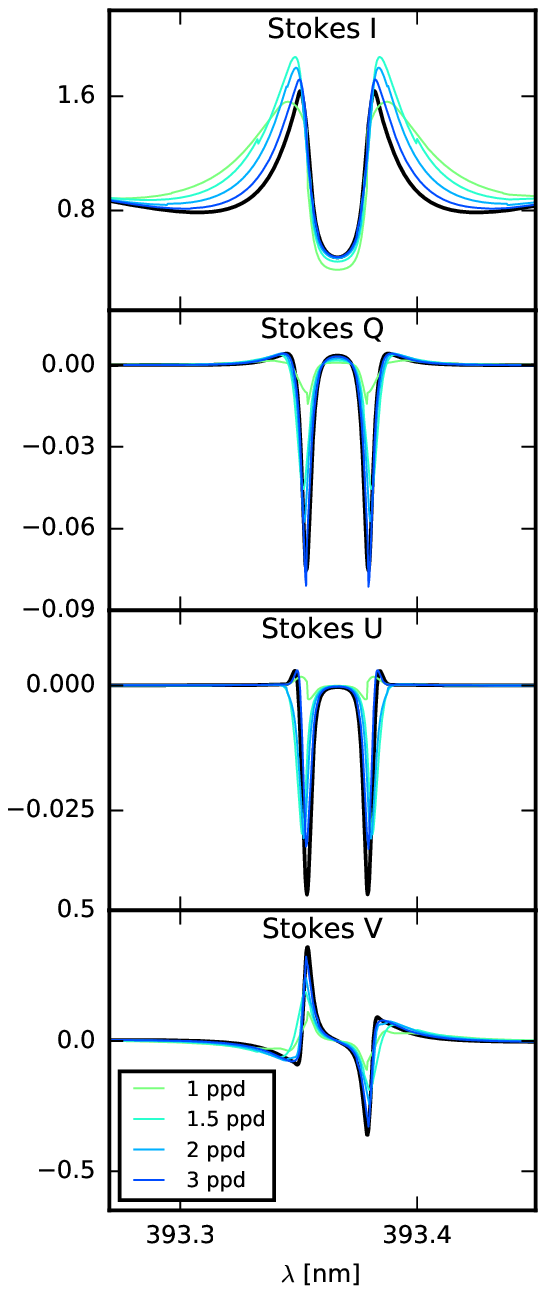}
\caption{Emerging Stokes profiles calculated with the second-order pragmatic method for
the Sr~{\sc i} line at 4607.3 {\rm \AA} for the Bifrost-1 atmospheric model (first column);
the Ca~{\sc ii} line at 8542.1 {\rm \AA} for the CO$^5$BOLD-2 atmospheric model (second column);
and the Ca~{\sc ii} line at 3933.7 {\rm \AA} for the FALC atmospheric model (third column).
The atmospheric models are described in Figure~\ref{fig:models}.
The different colors indicate different grid-point densities according the coding in the inset at the bottom of the last column.
The reference emergent Stokes profile (solid black) is calculated using a sampling with $10^3$ ppd.}
\label{fig:sri_pr}
\end{figure*}

Figure~\ref{fig:sri_pr} shows the emerging Stokes profiles,
synthesized with the second-order pragmatic method for three different poorly-sampled atmospheric models.
Figures~\ref{fig:sri}-\ref{fig:mgii} give the log-log representation
of the global error in the emergent Stokes vector profiles as a function of the number of ppd of continuum optical depth.
It would certainly be incautious to draw drastic conclusions on the basis of a qualitative comparison of small details of the error curves.
However, the numerical tests allow for different general considerations. The numerical results are listed in the following according the formation regions of the spectral lines.
Note that there is a limit (usually for  $\ge50$ ppd) beyond which the accuracy of fourth-order methods
is limited by machine precision and the fourth-order convergence may not be respected anymore.
\subsection{Photospheric spectral lines}\label{subsec:5.1}
This section analyzes the numerical synthesis of the photospheric Sr~{\sc i} line at 4607.3 {\rm \AA} and Fe~{\sc i} lines at 6301.5 and 6302.5 {\rm \AA}.
% \citet{bianda2014} use the Sr~{\sc i} 4607.3 {\rm \AA} spectral line in the synoptic program devoted
% to explore the nature of the turbulent magnetic field and the possible role of a local dynamo for the generation of these fields.
\citet{bianda2018} use the Sr~{\sc i} 4607.3 {\rm \AA} spectral line to infer
the spatial and temporal variation of the weak subgranular magnetic field with the Hanle effect.
Moreover, \citet{alsinaballester2017} studied this spectral line
considering PRD phenomena in the general Hanle-Zeeman regime.
%Besides that, the same numerical analysis is performed for the Fe~{\sc i} lines at 6301.5 and 6302.5 {\rm \AA}. 
The Fe~{\sc i} spectral lines at 6301.5 and 6302.5 {\rm \AA}  are often used to determine the magnetic field vector in the solar photosphere,
and they are the standard lines of the Advanced Spectropolarimeter \citep{skumanich1994} and of the Hinode Spectropolarimeter \citep{tsuneta2008}. 
%These lines are also used for the simultaneous inversion of spectral line pairs \citep[e.g.,][]{orozco_suarez2010}.
Moreover, \citet{bellot_rubio+al1998} and, lately, \citet{delacruz_rodriguez+piskunov2013} already used these spectral lines to analyze the performances of different formal solvers.

Here, these photospheric lines are synthesized under the assumption of LTE conditions.
% 
% No essential difference in the error curves is noticed between the FAL and the R-MHD atmospheric models, i.e., in the comparison of the first and second row of Figures~\ref{fig:sri} and~\ref{fig:fei}. In fact, the FALC, the Bifrost, and the CO$^5$BOLD atmospheric models do not differ very much in the photosphere.
% The comparison of Figures~\ref{fig:sri} and~\ref{fig:fei} does not reveal essential any difference in the error curves
% between the Sr~{\sc i} line at 4607.3 {\rm \AA} and the Fe~{\sc i} lines at 6301.5 and 6302.5 {\rm \AA}.
% Moreover, t
The comparison of the first and second row of Figure~\ref{fig:sri} does not reveal any essential difference between the error curves
of the FAL and the Bifrost atmospheric models.
In fact, the behavior of these discrete models do not differ very much in the photosphere,
while Figure~\ref{fig:fei} indicates that the higher intermittency of the CO$^5$BOLD-1 model weakens the performance of high-order methods.

It is very difficult to judge and compare the different formal solvers in the pre-asymptotic regime (i.e., where the convergence rate of numerical schemes may be not respected). 
Indeed, insufficient grid-point densities do not allow for high-order convergence and the accuracy of numerical schemes strongly depends on the specific sampling.
%, where the erratic pre-asymptotic behavior of the error curve has been frozen out. 
In any case, errors in the pre-asymptotic regime are usually $\ge10^{-1}$ and the synthesized Stokes profiles cannot be considered accurate.
The first column of Figure~\ref{fig:sri_pr} gives a glimpse on the stable pre-asymptotic behavior of the second-order pragmatic method in
the synthesis of the Sr~{\sc i} line at 4607.3 {\rm \AA} with very coarse Bifrost-1 grids.

% In fact, the DELO approach is thought to remove stiffness from the problem \citep{janett2017a},
% but it usually does not improve accuracy in the asymptotic regime.
%The asymptotic higher convergence of high-order schemes starts to be relevant and they 
Asymptotic convergence rates become relevant above 3 ppd.
In this regime, high-order schemes outperform low-order methods:
fourth-order methods require among 10-15 ppd to achieve an accuracy of $E_i\le10^{-4}$, for $i=1,2,3,4$ (see Equation~\eqref{error}).
Third-order methods need about twice as many ppd to reach the same accuracy,
whereas second-order methods require prohibitive numerical grids to produce such highly accurate spectra.

In the asymptotic regime, the accuracy and the order of convergence achieved by DELO methods 
and corresponding order non-DELO methods (with the exception of the pragmatic formal solvers) are almost identical and no clear differences are visible.
Second- and third-order pragmatic methods achieve the predicted order of accuracy.
\subsection{Lower chromospheric spectral lines}\label{subsec:5.2}
This section analyzes the numerical synthesis of the Ba~{\sc ii} line at 4554.0 {\rm \AA} and the central line of the Ca~{\sc ii} infrared triplet at 8542.1 {\rm \AA}.
The high photospheric/low chromospheric Ba~{\sc ii} 4554.0 {\rm \AA} line drew attention because of the magnetic sensitivity
of the strong scattering polarization signal that it shows in the so-called Second Solar Spectrum \citep[e.g.,][]{stenflo1997,belluzzi2007,faurobert2009,ramelli2009,smitha2013}.
% This spectral lines is used to \textcolor{blue}{\dots(any ideas?)\dots} taking into account quantum interferences and partial redistribution \citep[e.g.,][]{belluzzi2011,smitha2014}. 
%Besides that, the same numerical calculations are performed for the Ca~{\sc ii} infrared line at 8542.1 {\rm \AA}.
The low chromospheric Ca~{\sc ii} 8542.1 {\rm \AA} line has been extensively used as diagnostics because of its high sensitivity to photospheric and chromospheric responses \citep{delacruz_rodriguez+piskunov2013,quintero2016}.
% Similarly to the numerical results for photospheric spectral lines, no relevant difference in the error curves is noticed between the FAL and the R-MHD vertical columns cases, i.e., in the comparison of first and second rows of Figures~\ref{fig:baii} and~\ref{fig:caii}.

Here, NLTE effects are included for both spectral lines in terms of the field-free approximation. 
The comparison of the first and second row of Figures~\ref{fig:baii} and~\ref{fig:caii} shows relevant differences between the error curves
of FAL and R-MHD atmospheric models.
The intermittency of R-MHD single-pixel models in the low chromosphere 
weakens the performance of high-order methods.
In fact, the break-even point where high-order methods start to outperform
low-order methods depends on the smoothness of the atmospheric model:
it is at about 5 ppd when using FAL models and at 10-15 ppd for the Bifrost-2 model.
The strongly intermittent behavior in the outer layers of the CO$^5$BOLD atmospheric models
thwarts high-order convergence below 20-30 ppd, defeating the purpose of using high-order methods.
For a given accuracy, the synthesis of chromospheric spectral lines generally requires a slightly higher density of grid points
than the synthesis of photospheric spectral lines.
This is also due to the chosen homogeneous sampling in $\log \tau_c$, which is more suitable
for the synthesis of spectral lines forming around $\tau_c=1$ than to spectral lines forming in higher layers of the atmosphere.

The DELO-linear method usually maintains the pre-asymptotic error smaller than the trapezoidal method.
This is due to the accurate representation of the fast exponential attenuation of optically thick cells guaranteed by its $L$-stability. 
The occurrence of numerical instability of high-order methods is evident:
the Adams-Moulton 3 method shows strong instabilities when treating optically thick cells, because of its limited stability region.
Moreover, the global error of both Adams-Moulton 3 and DELO-parabolic methods grows, instead of decreasing,
as the density of grid points increases.
Additional numerical test calculations have excluded that round-off errors are the cause of this kind of instability, which instead
might be due to the compound effect of variable cell size and variable propagation matrix coefficients,
that modifies the stability function of multistep methods.
The second column of Figure~\ref{fig:sri_pr} shows
the synthesis of the Ca~{\sc ii} 8542.1 {\rm \AA} performed by the second-order pragmatic method with very coarse CO$^5$BOLD-2 grids.
It demonstrates that this method is stable even in the case of intermittent poorly-sampled atmospheric models.

The accuracy and the order of convergence achieved by DELO methods
and corresponding order non-DELO methods are almost identical in the asymptotic regime.
Similarly to the case of photospheric lines, the second- and third-order pragmatic methods do not suffer from numerical instability and achieve the predicted order of accuracy.
In some cases, they even achieve higher accuracy with respect to corresponding order schemes.
\vspace*{0.3cm}
\subsection{Upper chromospheric spectral lines}\label{subsec:5.3}
This section analyzes the numerical synthesis of the Ca~{\sc ii} K line at 3933.7 {\rm \AA} and the Mg~{\sc ii} k line at 2795.5 {\rm \AA}.
The Ca~{\sc ii} K line is the strongest spectral line in the visible solar spectrum,
shows very extended wings, and forms across a large range of atmospheric heights.
%Its correct modeling requires effects of partial frequency redistribution to be taken into account.
The core region is of particular interest because it carries information on the physical conditions of the high chromosphere
\citep[e.g.,][]{solanki1991,carlsson1997,solanki2004,anusha2013,bjorgen2018}.
% This spectral line is particularly appropriate for chromospheric diagnostics \textcolor{blue}{(why?)} and is influenced by partial frequency redistribution scattering mechanism and line polarization through the Hanle effect \citep[e.g.,][]{anusha2013,anusha2015}.
The Mg~{\sc ii} k line is an important diagnostics of 
the Interface Region Imaging Spectrograph (IRIS) space telescope \citep{depontieu2014}.
Moreover, \citet{leenaarts2013a,leenaarts2013b} synthesized the intensity profile of this spectral line from 3D R-MHD solar atmospheric models considering PRD.
Scattering polarization in this line has been modeled with a FALC atmosphere accounting for PRD effects by \citet{belluzzi2012} and, subsequently,
\citet{alsinaballester2016} also took the joint action of the Hanle and Zeeman effects into account.
%Particular interest towards this spectral line is given by 
The new instrument ``Chromospheric LAyer SpectroPolarimeter''
drew additional interest toward this spectral line \citep[CLASP2,][]{narukage2016}.

Here, NLTE effects are included for both spectral lines in terms of the field-free approximation.
Figures~\ref{fig:caii2} and~\ref{fig:mgii} show that high-order methods are well-performing for the case of the smooth FAL atmospheric models only.
%CO$^5$BOLD atmospheric models lack a proper transition region from the chromosphere to the corona, where the core of Mg~{\sc ii} k line forms.
Numerical methods perform very inefficiently in the absence of sufficient local smoothness
because discontinuities (or sharp gradients) might drastically increase local errors,
thwarting high-order convergence \citep[e.g.,][]{dieci2012}.
In practice, the high intermittency of the Bifrost and CO$^5$BOLD single-column models
in the chromospheric layers leads to an order of convergence ``breakdown'',
% \footnote{The absence of sufficient local smoothness in the problem
% introduce significant local errors in the numerical solution, which might even reduce the convergence order of the method.},
making the application of high-order schemes pointless \citep[e.g.,][]{mannshardt1978}.

The CO$^5$BOLD-2 atmosphere used for the convergence plots of Figure~\ref{fig:mgii} (bottom row) has
no proper chromospheric temperature rise. Correspondingly, the core of the Mg~{\sc ii} k
line has no k$_2$ emission peaks in this case. This may not seem realistic bearing
non-resolved observed profiles with clear k$_2$v and k$_2$r emission peaks and a
central k$_3$ core in mind. However, single columns of 3D R-MHD models
show a large variety of temperature profiles, and \citet{leenaarts2013b} report
that 0.1\% of the Mg~{\sc ii} h and k profiles from a Bifrost model atmosphere did not have
any emission peaks, although normally the Bifrost model features a proper chromospheric
temperature rise and transition zone. For this reason, this seemingly unrealistic example is included in the numerical tests.

Once again, the DELO-linear method usually maintains  the pre-asymptotic error equal or smaller than
the corresponding error of the trapezoidal method.
Adams-Moulton 3 and DELO-parabolic methods suffer from the same instability issues already encountered with the synthesis of low chromospheric lines. 
The third column of Figure~\ref{fig:sri_pr} gives a glimpse on the pre-asymptotic stable behavior of the second-order pragmatic method in
the synthesis of the Ca~{\sc ii} K line with very coarse FALC grids.

As for the previous cases, the accuracy and the order of convergence achieved by DELO methods
and corresponding order non-DELO methods are almost identical in the asymptotic regime.
As before, second- and third-order pragmatic methods do not suffer from numerical instability, achieving the predicted order of accuracy.

Note that the homogeneous sampling in $\log \tau_c$ might not be suitable for analyzing the synthesis of Stokes profiles of upper chromospheric spectral lines,
whose cores are forming far away from the formation region of the continuum.
%This fact hinders the numerical analysis previously performed for photospheric and lower chromospheric spectral lines.
However, alternative tests on atmospheric models re-sampled homogeneously in $\log\tau_\ell$ (where $\tau_\ell$ indicates the line center optical depth) show similar results.
\section{Conclusions}\label{sec:sec6}
% %%%%%%%%%%%%%%%%%%%%%%%%%%%%%%%%%%%%%%%%
This paper investigates different actors that play a relevant role in the numerical integration of the polarized radiative transfer equation:
in particular, the numerical scheme, the atmospheric model (and its sampling), and the spectral line.
Special attention is paid to assess the mathematical predictions given by \citet{janett2017a,janett2017b} and~\citet{janett2018a},
who characterized different formal solvers (in terms of order of accuracy, numerical stability, and computational cost)
and investigated the stiffness of Equation~\eqref{eq:RTE}.
Although NLTE effects are taken into account in the computation of the considered spectral lines, 
the analysis is limited to the numerical integration of Equation~\eqref{eq:RTE} and does not concern the full NLTE iterative problem.
%The numerical synthesis of Stokes profiles is analyzed for different formal solvers with different spectral lines and different realistic atmospheric models.

The urgency of high-order formal solvers has been often promoted,
e.g., in LTE inversions \citep{bellot_rubio+al1998} and in full NLTE problems \citep{trujillo_bueno2003}.
%In the radiative transfer of polarized light, the high-order convergence of a numerical method is, however, only relevant when the pre-asymptotic behavior of the error curve is avoided.
On the other hand, the numerical tests in Section~\ref{sec:sec5} indicate that the use of high-order methods is befitting
only if the relevant layers of the atmospheric model guarantee sufficient smoothness in the solution
(i.e., when the pre-asymptotic behavior of the error curve is avoided).
In the case of FAL models, high-order schemes generally outperform low-order methods with a density of grid points larger than 3-5 ppd.
Meanwhile, the high intermittency of 3D R-MHD models might thwart high-order convergence even in fine numerical grids,
making the application of high-order schemes pointless (or even noxious).
In the asymptotic regime, the accuracy achieved by DELO methods and corresponding order non-DELO methods is almost identical.

% that the predicted orders of accuracy of the different formal solvers are confirmed even when dealing with semi-empirical and R-MHD atmospheric models. In the asymptotic regime, the order of accuracy of the different methods is maintained, as shown by Figures~\ref{fig:sri}-\ref{fig:caii}. In the asymptotic regime, the accuracy achieved by DELO methods and corresponding order non-DELO methods is almost identical.

The numerical results confirm the predictions given by \citet{janett2018a}.
First, Figure~\ref{fig:conversion} shows that the conversion to optical depth effectively removes stiffness from Equation~\eqref{eq:RTE}
because it attenuates variation of the propagation matrix elements along the integration path.
Second, $L$-stability guarantees the exponential attenuations of the radiation across optically thick cells, avoiding numerical oscillations.
Consequently, the $L$-stable DELO-linear method always keeps the error $E_i\le1$ (for $i=1,2,3,4$)
and it is, consequently, particularly suitable to coarse atmospheric models with optically thick cells.
Third, the Adams-Moulton 3 method shows strong instabilities when treating optically thick cells because of its limited stability region.
In addition, both Adams-Moulton 3 and DELO-parabolic methods occasionally present
a particular kind of numerical instability, where
the global error grows as the density of grid points increases.
The origin of this kind of instability might be the compound effect of variable cell size and variable propagation matrix coefficients,
which modify the stability function of multistep methods.
For this reason, multistep schemes are not recommended for practical applications.
Furthermore, the B\'ezier strategy (intended to avoid overshooting in the absorption and emission coefficients) does not provide better results when treating realistic and intermittent atmospheric models. This is shown by the second row of Figures~\ref{fig:sri}-\ref{fig:mgii}, 
which expose the congruence between cubic B\'ezier method and the cubic Hermitian method.

The choice of the optimal formal solver definitely depends on the particular problem,
when opting for the cheapest numerical scheme that satisfies accuracy requirements (and ensures numerical stability).
In conclusion, if the atmospheric model lacks sufficient local smoothness or if a relatively large error is accepted (say $E_i\approx10^{-2}$),
the second-order pragmatic method is recommended and the DELO-linear method still represents a solid alternative with an easier implementation.
If higher accuracy is necessary (say $E_i\le10^{-3}$) and smoothness in atmospheric models %sufficiently dense spatial grids are guaranteed,
is guaranteed,
the third-order pragmatic method is recommended.

The performance of the numerical methods on discontinuous atmospheric models
requires a deeper investigation.
However, the structure of pragmatic methods might also be suitable to switch to suitable techniques,
when facing discontinuities in the integration interval.

%This indicates that standard samplings (e.g., FAL models with 70 grid points) could be unsuitable, if trying to achieve highly accurate results, i.e., $E_i\le10^{-4}$.
%
\acknowledgments
The financial support by the Swiss National Science Foundation (SNSF) through grant ID 200021\_159206 is gratefully acknowledged.
Special thanks are extended to E. S. Carlin and F. Calvo for providing the 3D R-MHD Bifrost and CO$^5$BOLD atmospheric models, respectively, and to E. Alsina Ballester  
for reading and commenting on a previous version of the paper.
\appendix
\section{Pragmatic methods pseudocodes}\label{appendix:B}
 \vspace*{0.5cm}
\begin{algorithm}[H]\label{alg1}
 %\KwData{this text}
 %\KwResult{how to write algorithm with \LaTeX2e }
 %initialization\;
 \For{$i=1$ to $N$}{
  compute $\Delta \tau$, $\phi_{\text{\rm \tiny H}}$, and $\phi_{\text{\rm \tiny T}}$\;
  \uIf{$(\phi_{\text{\rm \tiny H}} < 1\;\lor\;\Delta \tau<\tau_1)$}{
   
   $\Rightarrow$ Heun's method ($\mathbf{\Psi}^E$)\;
   
   }
   \uElseIf{$(\phi_{\text{\rm \tiny T}} < 1\;\land\;\Delta \tau<\tau_2)$}{
     
    $\Rightarrow$ Trapezoidal method ($\mathbf{\Psi}^A$)\;
    
  }\Else{$\Rightarrow$ DELO-linear method ($\mathbf{\Psi}^L$)\;
  
 }
 }\caption{Second-order pragmatic method.
 The choice of free parameters $\tau_1=10^{-3}$ and $\tau_2=7$ has proven good results.}
 \end{algorithm}
 \vspace*{0.5cm}
 \begin{algorithm}[H]\label{alg2}

  \For{$i=1$ to $N$}{
  compute $\Delta \tau$, $\phi_{\text{\rm \tiny RK3}}$, and $\phi_{\text{\rm \tiny H3}}$\;
  \uIf{$(\phi_{\text{\rm \tiny RK3}} < 1\;\lor\;\Delta \tau<\tau_1)$}{
   
   $\Rightarrow$ Runge-Kutta 3 method ($\mathbf{\Psi}^E$)\;
   
   }
   \uElseIf{$(\phi_{\text{\rm \tiny H3}} < 1\;\land\;\Delta \tau<\tau_2)$}{
     
    $\Rightarrow$ Hermitian 3 method ($\mathbf{\Psi}^A$)\;
    
  }\Else{$\Rightarrow$ DELO-linear method ($\mathbf{\Psi}^L$)\;
  
 }
 }\caption{Third-order pragmatic method. 
 The choice of free parameters $\tau_1=10^{-3}$ and $\tau_2=10$ has proven good results.}

 \end{algorithm}
\section{Error calculation}\label{appendix:A}
Denoting with $\mathbf I^{\rm ref}(\nu)$ and $\mathbf I^{\rm num}(\nu)$ the reference and the numerically computed emergent Stokes vectors, respectively, at the frequency $\nu$, the global error for each Stokes vector component is computed as 
\begin{equation}
E_i=\frac{\displaystyle \max_{\nu}| I_i^{\rm ref}(\nu)- I_i^{\rm num}(\nu)|}{{\displaystyle \max_{\nu}} \; I_i^{\rm ref}(\nu)-{\displaystyle\min_{\nu}} \; I_i^{\rm ref}(\nu)}\,,
\text{ for }i=1,2,3,4\,,
\label{error}
\end{equation}
where $i=1,2,3,4$ indicate the four Stokes parameters $(I,Q,U,V)$. The error is given by the maximal discrepancy between the reference and the simulated Stokes parameter over the spectral interval considered, normalized by the maximal amplitude in the reference profile. The reference emergent Stokes profile is calculated with the cubic Hermitian method (and cross-checked with the quadratic DELO-B\'ezier method) using a hyperfine grid sampling with $10^3$ ppd of continuum optical depth.
The infinite norm used for the global error is sensitive to outliers at single wavelengths and error curves may show non-monotonic behavior, especially in pre-asymptotic regime.

%%increasing the accuracy of more than one order of magnitude.
%An explicit example is shown in Figure~\ref{fig:profile}.
%% The following command ends your manuscript. LaTeX will ignore any text
%% that appears after it.
\bibliographystyle{apj}
\bibliography{bibfile4}
\begin{figure*}
\centering
\includegraphics[width=1.\textwidth]{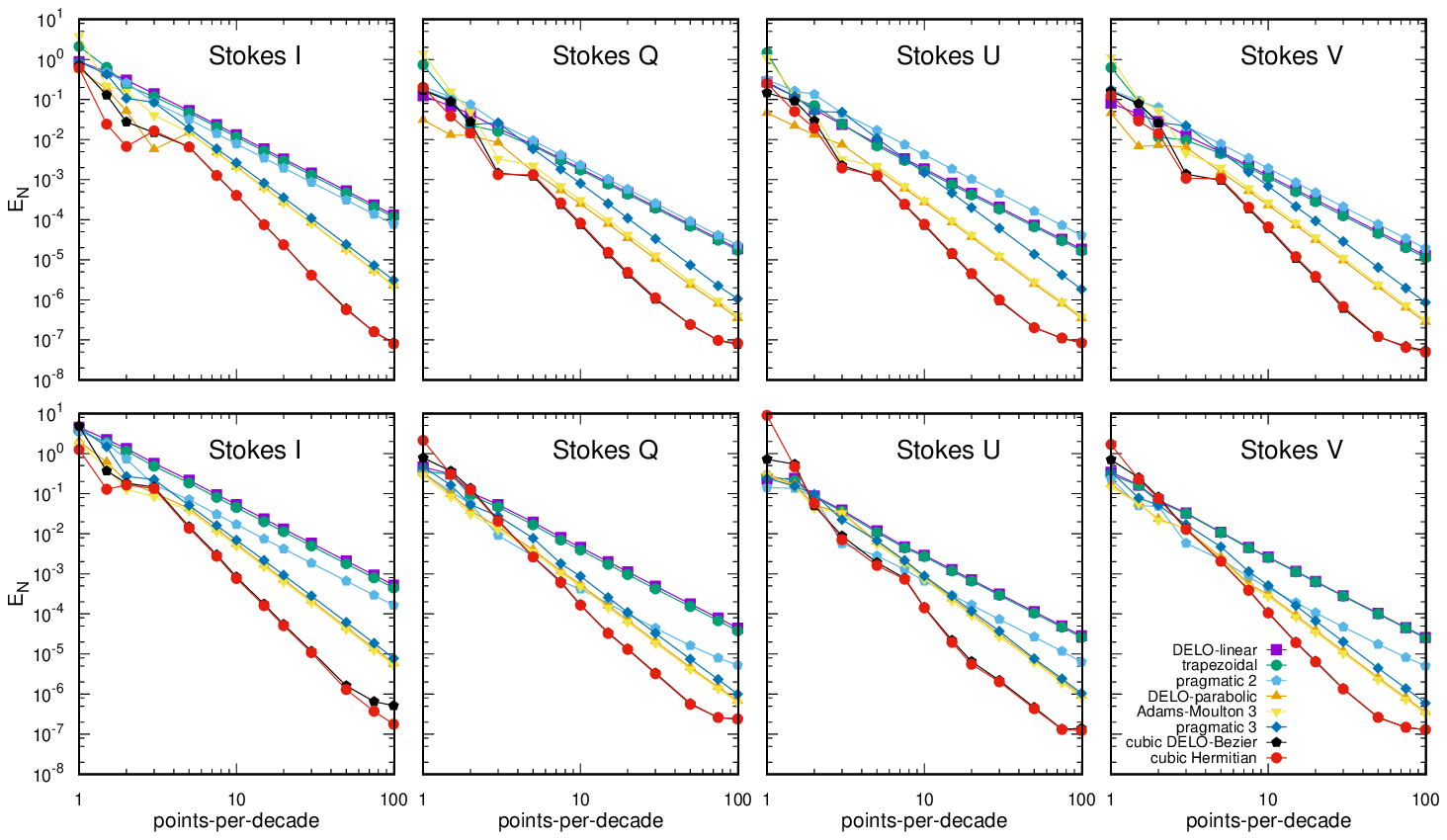}
  \caption{The log-log representation of the global error for the Stokes vector components $I,Q,U$, and $V$ as a function of the number of ppd of continuum optical depth for the formal solvers (color-coded) presented in Section~\ref{sec:sec3}. The Sr~{\sc i} line at 4607.3 {\rm \AA} is considered for the FALC atmospheric model (first row) and the Bifrost-1 atmospheric model (second row). The atmospheric models are described in Figure~\ref{fig:models}. The global error is computed as exposed in Appendix~\ref{appendix:A}.}
  \label{fig:sri}
\centering
\includegraphics[width=1.\textwidth]{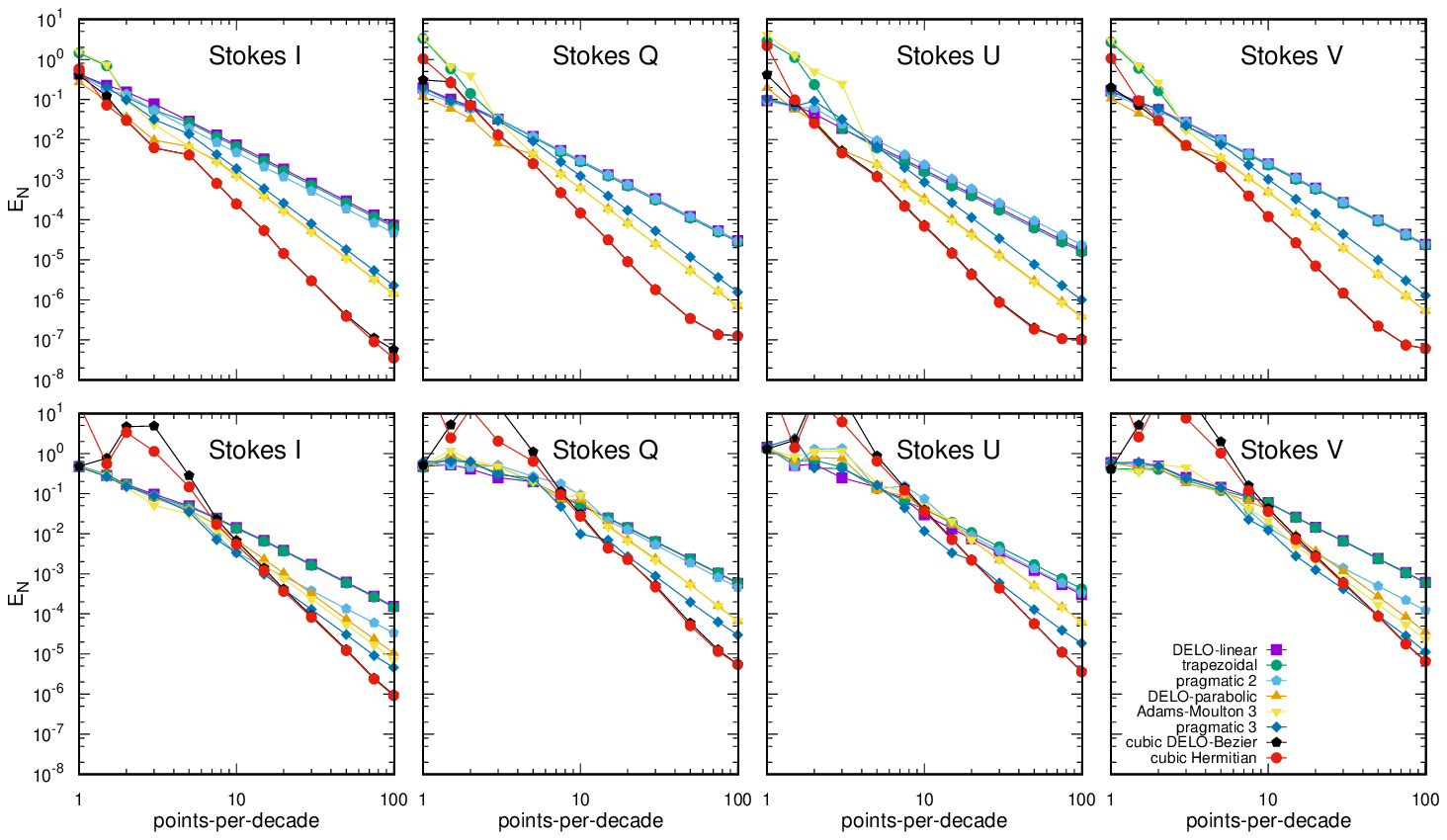}
  \caption{Same as Figure~\ref{fig:sri}, but for the Fe~{\sc i} lines at 6301.5 and 6302.5 {\rm \AA}, respectively, and for the FALF atmospheric model (first row) and the CO$^5$BOLD-1 atmospheric model (second row).}
  \label{fig:fei}
\end{figure*}
\begin{figure*}
\centering
\includegraphics[width=1.\textwidth]{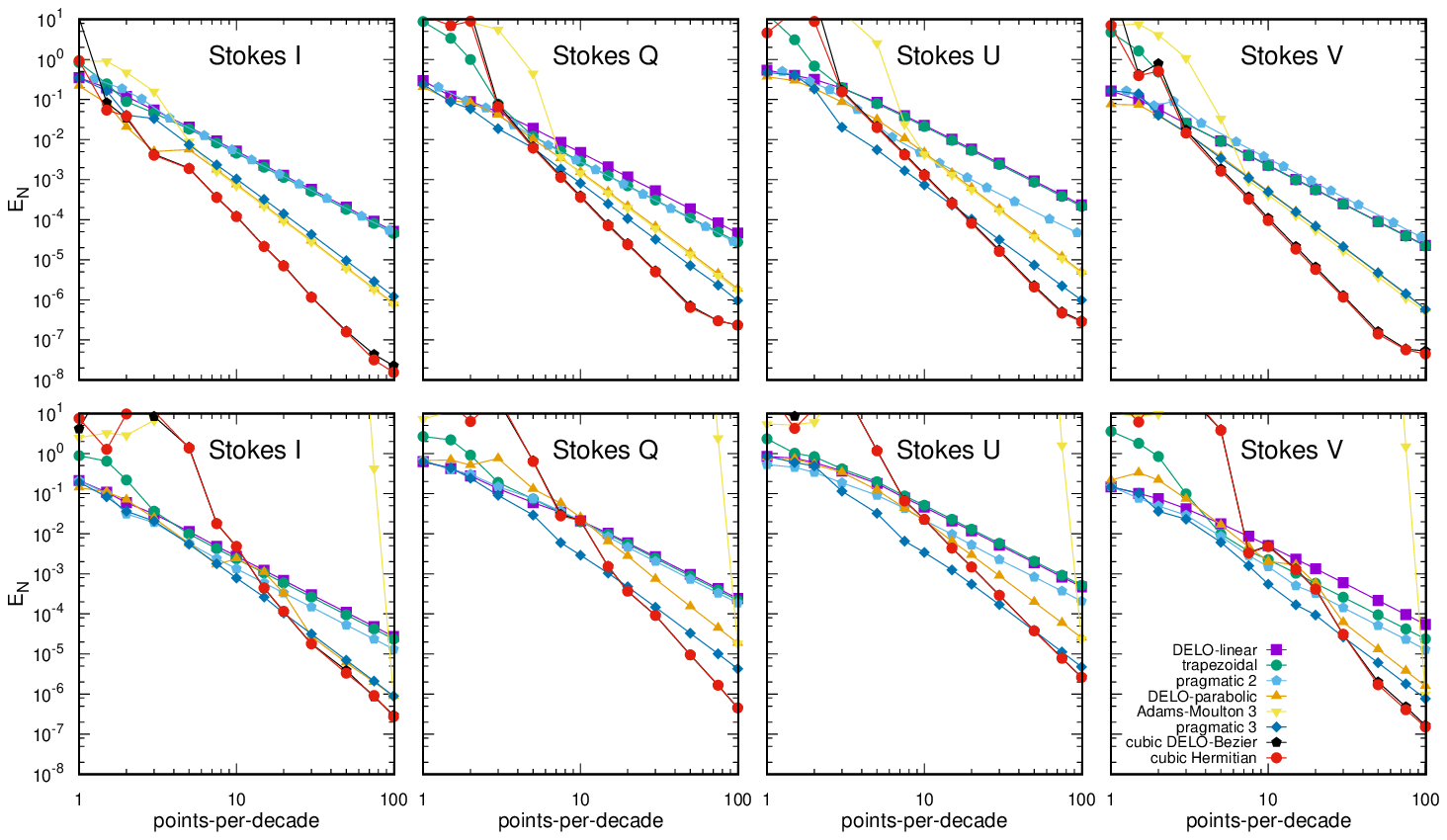}
  \caption{Same as Figure~\ref{fig:sri}, but for the Ba~{\sc ii} line at 4554.0 {\rm \AA} and for the FALC atmospheric model (first row) and the Bifrost-2 atmospheric model (second row).}
  \label{fig:baii}
\centering
\includegraphics[width=1.\textwidth]{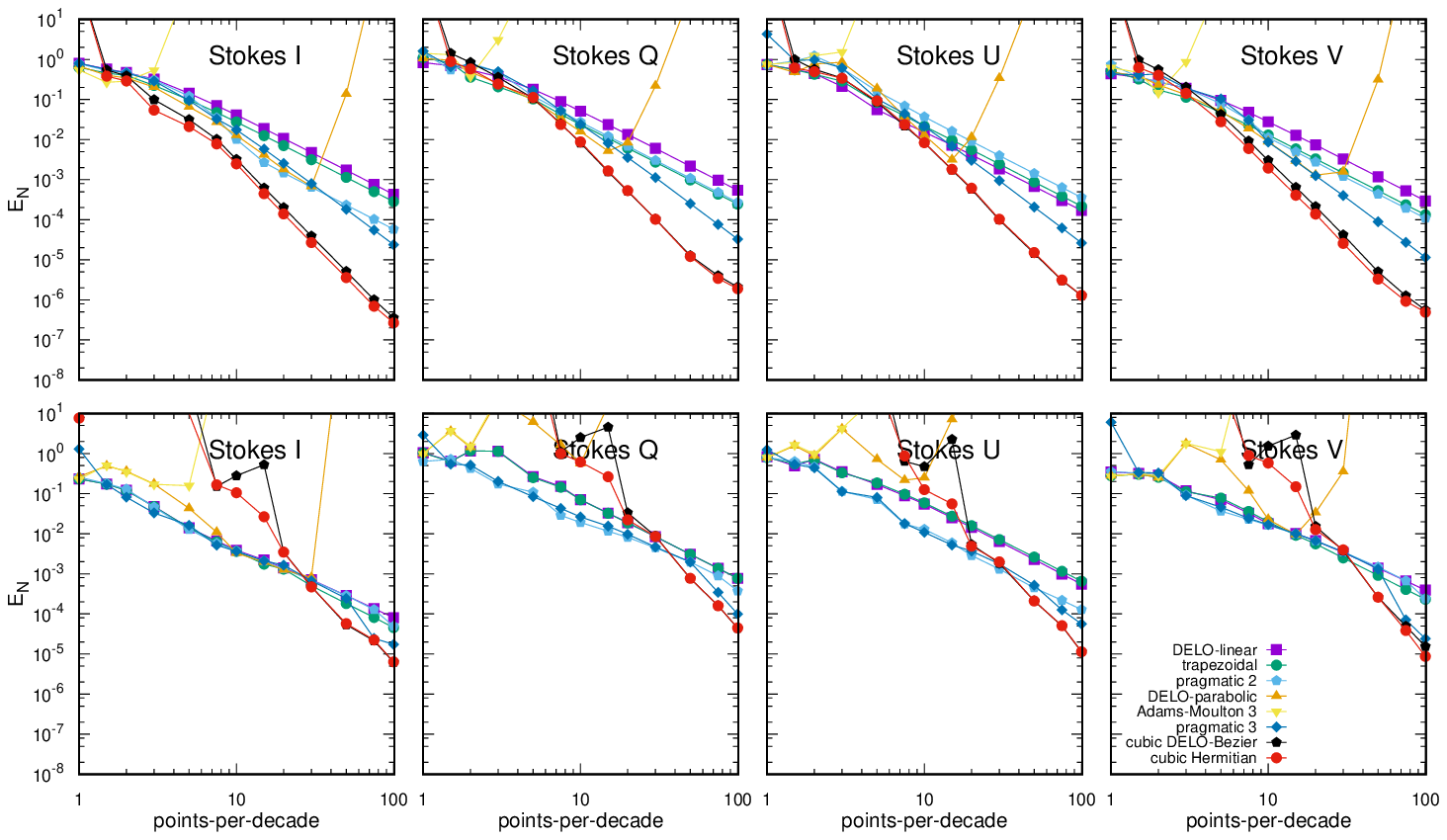}
  \caption{Same as Figure~\ref{fig:sri}, but for the infrared Ca~{\sc ii} line at 8542.1 {\rm \AA} and for the FALF atmospheric model (first row) and the CO$^5$BOLD-2 atmospheric model (second row).}
  \label{fig:caii}
\end{figure*}
\begin{figure*}
\centering
\includegraphics[width=1.\textwidth]{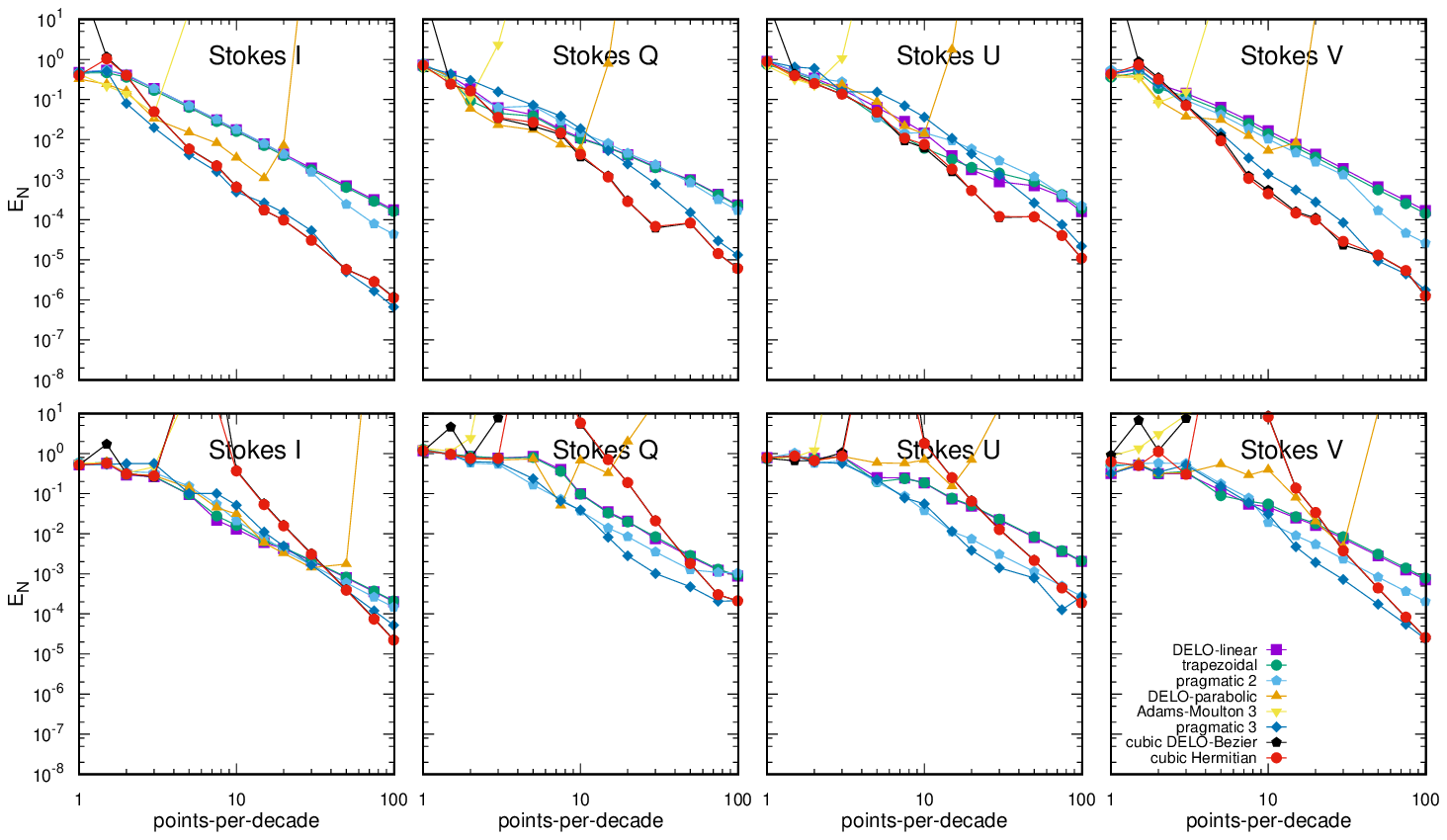}
  \caption{Same as Figure~\ref{fig:sri}, but for the Ca~{\sc ii} K line at 3933.7 {\rm \AA} and for the FALC atmospheric model (first row) and the Bifrost-2 atmospheric model (second row).}
  \label{fig:caii2}
\centering
\includegraphics[width=1.\textwidth]{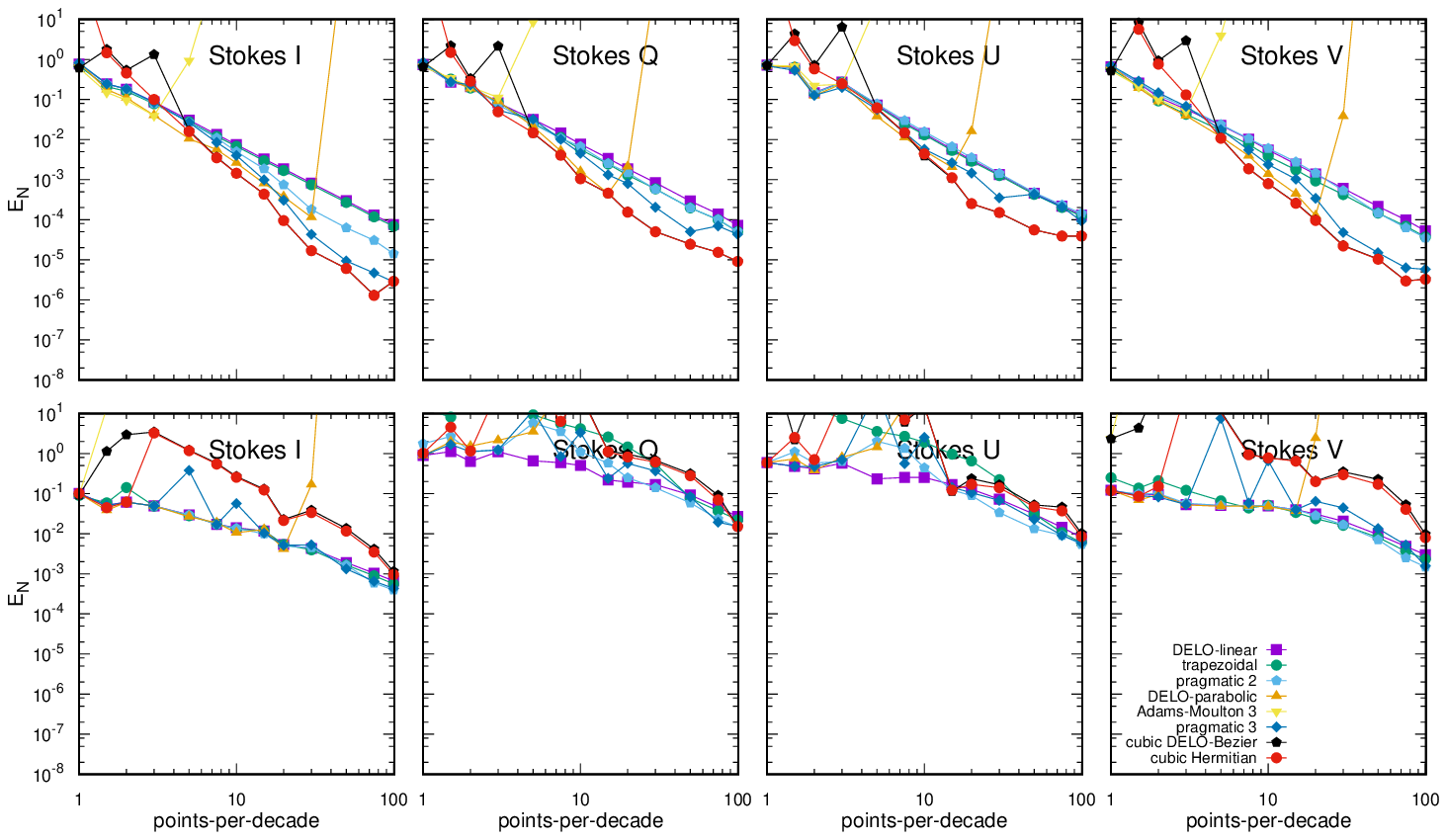}
  \caption{Same as Figure~\ref{fig:sri}, but for the Mg~{\sc ii} k line at 2795.5 {\rm \AA} and for the FALF atmospheric model (first row) and the CO$^5$BOLD-2 atmospheric model (second row).}
  \label{fig:mgii}
\end{figure*}
\end{document}